\documentclass[a4paper,11pt]{article}
\pdfoutput=1 

\usepackage{jheppub} 

\usepackage[english]{babel}
\usepackage{graphicx,epsfig}
\usepackage{mathrsfs,amsmath}
\usepackage{slashed}

\usepackage{tikz-feynman}

\usepackage{color}

\usepackage{enumerate}
\usepackage{epsfig}
\usepackage{soul}
\usepackage{verbatim,setspace}
\usepackage{xparse}
\usepackage{tikz}
\usetikzlibrary{calc}
\usepackage{floatrow}
\usepackage{colortbl}
\usepackage{arydshln}
\usepackage{hhline}
\usepackage{pifont}
\newcommand{\cmark}{\ding{51}}%
\usepackage{floatrow}
\usepackage{subcaption}

\newcommand{\gre}{\cellcolor{green!100}}
\newcommand{\yel}{\cellcolor{yellow!100}}
\newcommand{\ora}{\cellcolor{orange!100}}
\newcommand{\redd}{\cellcolor{red!100}}

\newcommand{\cL}{{\cal L}}

\newcommand{\be}{\begin{equation}}
\newcommand{\ee}{\end{equation}}
\newcommand{\bea}{\begin{eqnarray}}
\newcommand{\eea}{\end{eqnarray}}

\DeclareMathSymbol{\mg}{\mathrel}{symbols}{"1D}

\newcommand{\MET}{E\llap{/\kern1.5pt}_T}
\newcommand{\nn}{\nonumber}

\newcommand{\df}{\dfrac}

\DeclareDocumentCommand{\hcancel}{mO{0pt}O{0pt}O{0pt}O{0pt}}{%
    \tikz[baseline=(tocancel.base)]{
        \node[inner sep=0pt,outer sep=0pt] (tocancel) {#1};
        \draw[black] ($(tocancel.south west)+(#2,#3)$) -- ($(tocancel.north east)+(#4,#5)$);
    }
}

\title{\boldmath  
On Minimal Dark Matter coupled to the Higgs}

\author[a,b]{Laura Lopez Honorez,}
\author[b]{Michel H.G. Tytgat,}
\author[a]{Pantelis Tziveloglou,}
\author[c]{Bryan Zaldivar}

\affiliation[a]{Theoretische Natuurkunde, Vrije Universiteit Brussel and The International Solvay Institutes, Pleinlaan 2, B-1050 Brussels, Belgium.}
\affiliation[b]{Service de Physique Th\'eorique, Universit\'e Libre de Bruxelles, C.P. 225, B-1050
Brussels, Belgium}
\affiliation[c]{LAPTh, Université de Savoie Mont Blanc
CNRS, B.P.110, F-74941 Annecy-le-Vieux, France}

\emailAdd{llopezho@ulb.ac.be}
\emailAdd{mtytgat@ulb.ac.be}
\emailAdd{pantelis.tziveloglou@vub.ac.be}
\emailAdd{bryan.zaldivar@lapth.cnrs.fr}

\abstract{ We provide a unified presentation of extensions of the Minimal Dark
Matter framework in which new fermionic electroweak multiplets are
coupled to each other via the Standard Model Higgs doublet. We study
systematically the generic features of all the possibilities, starting
with a singlet and two doublets (akin to Bino-Higgsino dark matter) up
to a Majorana quintuplet coupled to two Weyl quadruplets. We pay
special attention to this last case, since it has not yet been
discussed in the literature. We estimate the parameter space for
viable dark matter candidates. This includes an estimate for the mass
of a quasi-pure quadruplet dark matter candidate taking into account the Sommerfeld
effect. We also argue how the coupling to the Higgs can bring the
Minimal Dark Matter scenario within the reach of present and future
direct detection experiments.}


\preprint{ULB-TH/17-21, LAPTH-058/17}

\begin{document} 
\maketitle

\section{Introduction}
\label{sec:intro}

The Minimal Dark Matter (MDM)~\cite{Cirelli:2005uq,Cirelli:2009uv}
scenario is one of the simplest extensions of the Standard Model with a
dark matter (DM) candidate. It requires the addition of one single
(real or complex) scalar or (Majorana or Dirac) fermionic $SU(2)_L$
multiplet, with mass $M$ as the only free parameter. A mass splitting
between the components of the multiplet arises as a loop correction
and it is a generic outcome that the lightest component is neutral and
thus a potential dark matter candidate. Such a candidate is a WIMP
(indeed a perfect WIMP, or WIMP archetype, as it has only electroweak
interactions) and matching with the observed DM abundance points to
a specific prediction for the mass $M$ of the thermal candidate,
different for each representation, but all in the TeV range. The
precise determination of this mass is however notoriously delicate
because of non-perturbative effects that must be taken into account to
calculate the effective annihilation cross section of the DM in the
early Universe, a point to
which we shall come back.  The classification of possible $SU(2)_L$
representations may be further restricted by requiring the absence of
a Landau pole, potentially up to the Planck scale. For a fermion
(scalar) candidate, the largest admissible representation is in a
quintuplet (respectively septuplet) of $SU(2)_L$. Interestingly the
stability of DM may be automatic in the case of a quintuplet (in the
sense that the lifetime of the DM candidate is naturally long, even
taking into account the possible contribution from effective
operators), without the need to impose an {\em ad hoc} discrete $Z_2$
symmetry \cite{Cirelli:2005uq} (a scalar septuplet however, despite
being in a large representation of $SU(2)_L$, may be unstable at
one-loop \cite{DiLuzio:2015oha}). For this reason, depending on the
context or on the authors, Minimal Dark Matter may refer to the
quintuplet candidate only, or the whole set of admissible electroweak
multiplets; we will adopt the latter
definition.\footnote{Alternatively, a discrete symmetry may be a
  remnant of a gauge symmetry \cite{Krauss:1988zc}. This is the case
  for so-called matter parity in the framework of $SO(10)$ Grand
  Unified Theory \cite{Kadastik:2009dj}. Table 2 of
  \cite{Nagata:2015dma} lists all $SO(10)$ representations up to $\bf
  210$ and $\bf 210'$ that contain a DM candidate. They encompass all
  the MDM candidates up to a fermionic $SU(2)_L$ quadruplet ($\bf 660$
  is the smallest $SO(10)$ representation that contains a fermionic
  quintuplet).}

Minimal Dark Matter candidates, like potentially any WIMP, may be
searched experimentally. Most relevant for MDM are constraints from
indirect and direct searches (assuming that MDM is the dominant form
of DM within a standard cosmological evolution). First, direct
detection limits exclude any MDM candidate with non-zero hypercharge
(hence a Dirac fermion or a complex scalar) due to scattering off
nucleons through $Z$ boson exchange. Now, sufficient 
mass spitting between the neutral components can help to alleviate
such constraints, see e.g.~\cite{DeSimone:2010tf}. This is what we
will assume when quoting doublet and quadruplet cross-sections below.
For a Majorana or real scalar candidate, a coupling to nucleons arises
at one-loop (with only a spin-independent (SI) contribution in the
scalar case), see e.g.~\cite{Cirelli:2005uq} for a first
estimation. The scattering cross-section of MDM off nucleons has
 been carefully revisited at NLO in ref.~\cite{Hisano:2015rsa} and,
for a fermion MDM-proton scattering, in a representation of dimension
$n$ and of hypercharge $Y$, one has:
\begin{equation}
  \sigma_{\rm SI}=\frac{4}{\pi}\mu^2 f_p^2\quad{\rm with}\quad f_p=(n^2-4Y^2-1) f^W_p+Y^2 f^Z_p
  \label{eq:si}
\end{equation}
with $f_p^W=2.9\, 10^{-10}$ GeV$^{-2}$ and $f_p^Z=-1.8 \,10^{-10}$
GeV$^{-2}$ and $\mu=m_{DM}m_p/(m_{DM}+m_p)$ is the reduced mass.  Such
estimation gives rise to lower cross-sections than originally
estimated and appear to be above the neutrino floor (except for the
doublet) and potentially marginally testable by the Xenon 1T
\cite{Aprile:2015uzo} experiment. In particular, from
eq.~(\ref{eq:si}), one gets $\sigma_{\rm SI} = 8.4 \times 10^{-50}$
cm$^2$ for a fermion doublet, with $(n,Y)=(2,1/2)$, $\sigma_{\rm SI} =
2.7 \times 10^{-47}$ cm$^2$ for a triplet (or 3-plet for short), with
$(n,Y)=(3,0)$, $\sigma_{\rm SI} = 1.6 \times 10^{-46}$ cm$^2$ for a
quadruplet (4-plet), with $(n,Y)=(4,1/2)$, and finally, $\sigma_{\rm
  SI} = 2.4 \times 10^{-46}$ cm$^2$ for a quintuplet (5-plet), with
$(n,Y)=(5,0)$.  Notice that in all cases, one can also compute the
spin-dependent scattering on nucleons. We have checked that, at
tree-level, the spin-dependent scattering cross-sections are
way beyond current DM searches limits (for a recent analysis,
see e.g.~\cite{Hill:2014yxa}).

Indirect detection limits on MDM candidates are also strong, at least
assuming an Einasto or Navarro-Frenk-White (NFW) profiles for the dark
matter distribution in the Galaxy. This is because of the Sommerfeld
effect, which typically enhances the annihilation cross section of
MDM candidates at small relative velocities, giving rise to strong
gamma-ray spectral features, see
\cite{Lefranc:2016fgn,Ovanesyan:2014fwa,Baumgart:2014saa} for the wino
case and \cite{Cirelli:2015bda,Garcia-Cely:2015dda} for the
quintuplet.\footnote{See also \cite{Lefranc:2016fgn} for an appraisal
  of current and future constraints, including from dwarf spheroidal
  galaxies.} Notice that on general grounds, dark matter bound state
formation~\cite{MarchRussell:2008tu,Ellis:2015vaaw,vonHarling:2014kha,Wise:2014jva,Liew:2016hqo}
could also affect the dark matter annihilation. It has been shown that
the latter effect is expected to be relevant for quintuplet dark
matter, while it is negligible in the case of
the triplet~\cite{Asadi:2016ybp,Mitridate:2017izz}. In general, the wino-like dark
matter appears now strongly disfavoured by current indirect detection
searches~\cite{Ovanesyan:2016vkk} while the quintuplet could be
tested by very near future HESS-II data release on searches
for gamma-ray lines from the 10 years Galactic Center
data~\cite{HESStalk}.

Because of the advent of these constraints, it may be timely to consider possible variations around the MDM framework,
which at the same time may lead to a broader range of possible DM
candidates. As mentioned above, a basic assumption of this framework is that
there is one and only one electroweak multiplet.  This, in particular,
precludes Yukawa coupling to the SM Higgs doublet for fermionic
candidates.\footnote{For scalar MDM candidates, quartic couplings to the
  Higgs are  allowed for any representation, a scenario that has been
  much studied in the literature, see {\em
    e.g.}~\cite{Hambye:2009pw}.} A natural yet simple variation on the MDM framework is to
consider simultaneously different multiplets, in particular fermionic
representations that differ by isospin $\Delta I = 1/2$, that allows
for ``integrating the Higgs portal to fermion
DM''~\cite{Freitas:2015hsa}. A familiar instance is the neutralino of
the Minimal Supersymmetric Standard Model (MSSM), which is generically
a mixture of bino/higgsino/wino complex. Recently, there have been
much studies of DM candidates from mixed (as compared to pure)
representations: singlet-doublet ($\sim$ bino-higgsino)
\cite{Mahbubani:2005pt,D'Eramo:2007ga,Enberg:2007rp,Cohen:2011ec,Cheung:2013dua,Calibbi:2015nha,Banerjee:2016hsk,Freitas:2015hsa}
(see also \cite{Yaguna:2015mva} for the case of a Dirac singlet),
doublet-triplet ($\sim$ higgsino-wino)
\cite{Beneke:2016jpw,Freitas:2015hsa,Dedes:2014hga} and
 triplet-quadruplet \cite{Tait:2016qbg,Banerjee:2016hsk}.

In the present work, we complete this panorama by adding to this list the case
of two Weyl 4plets coupled to a Majorana 5-plet (thus called $5_M4_D$),
while discussing in an unified manner the rest of the HMDM candidates. This may be of particular interest given the special status of
the fermionic 5-plet within the MDM framework, as alluded to above.\footnote{Notice that the stability of the MDM 5-plet is accidental and rests on the assumption that there are no other degrees of freedom below, say, a GUT scale. Indeed, its decay into SM degrees of freedom is driven by a dimension 6 operator, through the $L H H H^\ast \sim (5,0)$ combination of SM fields. In the same way, a Dirac 4-plet would involve a 5 dimensional operator, with $L H H^\ast \sim (4, -1/2)$. Such operator would lead to its rapid decay. Thus, if the $4_D$ is not at the GUT scale and couples to a $5_M$, the latter  is no longer protected from decay. Hence, in our framework, a discrete parity must be imposed on all the new fermionic multiplets.} The
scenarios that we consider rest on only 4 free parameters: 2 bare masses
(one Dirac mass, $m_D$, and one Majorana mass, $m_M$), and two Yukawa
couplings to the Higgs, $y_1$ and $y_2$ hence 3 extra parameters
compared to the pure MDM case (in the sequel, we will refer to pure,
i.e. \`a la MDM, and mixed states). Considering thermal candidates
leaves a 3-dimensional subspace of possible candidates to explore. The
goal of this paper is to illustrate that, due to the Yukawa coupling
to the Higgs, Higgs coupled MDM (denoted HMDM in what follows)
scenarios allow to enlarge the DM mass range of pure MDM scenarios in
a controlled way, and to argue that they could potentially allow to
evade current indirect detection constraints while providing the
opportunity to give rise to a signal in near future direct detection
facilities.

The structure of the paper is as follows. We begin this article
describing the general properties of HMDM in a unified framework and
analyze the properties of the mass spectra of both neutral and charged
states in Sec.~\ref{sec:higgs-coupl-minim}. We then discuss the viable
parameter space for a HMDM dark matter candidate taking into account
non perturbative corrections to the processes of (co)-annihilation
making use of the SU(2)$_L$ symmetric limit and discuss briefly the
prospects for dark matter detection in Sec.~\ref{sec:dark-matt-phen}.
We finally conclude in Sec.~\ref{sec:concl-prosp} and provide some
extra material in the appendix.

\section{Higgs coupled Minimal Dark Matter (HMDM)}
\label{sec:higgs-coupl-minim}

We consider left-handed Weyl fermions, $\psi$ and $\tilde
\psi$, in a $2n$-dimensional representation of $SU(2)_L$ with hypercharge
$Y_{\psi} =-Y_{\tilde \psi} = 1/2$ ({\em i.e.} an anomaly free,
vector-like fermion), together with a Majorana
fermion, $\chi$, (hence with $Y_\chi=0$) in a $2n \pm 1$ representation of
$SU(2)_L$. Going to 4-components notation, one can construct the
Dirac fermion $2n$-plet as $\Psi=(\psi,\epsilon \tilde \psi^\dag)$,
with ($\epsilon=i \sigma_2$ the anti-symmetric tensor of SU(2)) and
$X=(\chi,\epsilon \chi^\dag)$ the Majorana fermion. As mentioned in
the introduction, the fermions quantum numbers are chosen so that
these fields may have a Yukawa coupling to the SM Higgs and contain a
neutral particle. To ensure DM stability, we assume
that all fields of the dark sector are odd under a $Z_2$ symmetry,
while the Standard Model particles are even.

\begin{center}
\begin{tabular}{*{4}{|p{2.11cm}}|l}
\hhline{*{4}{|-}|~}
\centering M\textbackslash D &\centering 2 & \centering 4  &\centering 6 &
\\ \hhline{*{4}{|-}|~}
\centering 1  &\gre\centering \cmark\,\cite{Mahbubani:2005pt,D'Eramo:2007ga,Enberg:2007rp,Cohen:2011ec,Cheung:2013dua,Calibbi:2015nha,Banerjee:2016hsk,Freitas:2015hsa}  & \gre &\ora
\\ \hhline{*{4}{|-}|~}
\centering 3 & \gre\centering \cmark\,\cite{Beneke:2016jpw,Freitas:2015hsa,Dedes:2014hga}  &\yel\centering\cmark\,\cite{Tait:2016qbg,Banerjee:2016hsk} &\ora
\\ \hhline{*{4}{|-}|~}
\centering 5 & \gre&\yel\centering \cmark &\qquad\,\redd\cmark \\ \hhline{*{4}{|-}|~}
\centering 7 &\ora &\ora    &\qquad\,\redd\cmark\\ \hhline{*{4}{|-}|~}
\end{tabular}\label{modelspace}
\captionof{table}{The HMDM Model Space. Check marks correspond to pairs of Dirac (D) and Majorana (M) DM representations that can have a Yukawa coupling to the SM Higgs. The green cells are models with a Landau pole (LP) for $\alpha_2$ at $\Lambda_{\rm LP} \geq M_{Pl}$, while the yellow, orange and red cells correspond to $\Lambda_{\rm LP}$ in $[M_{Pl},10^{10} {\rm GeV}]$, $[10^{10}  {\rm GeV},10^{5}  {\rm GeV}]$ and $ <10^{5} $ GeV respectively.}
\end{center}

As in the usual MDM framework, we may require that the DM sector does
not drive electroweak couplings to a Landau pole at a too low energy  scale. This requirement
sets upper limits on the possible pairs of Dirac (noted $D$) and
Majorana (resp. $M$) $SU(2)_L$ representations that are stronger than for pure
MDM candidates. This leads to the results summarized in
Table~\ref{modelspace}, where we show the $D/M$ pairs with,
respectively, no Landau pole below $\Lambda_{\rm LP} = M_{\rm Pl}$
(green cells), $\Lambda_{\rm LP} = 10^{10}\,$GeV (yellow cells) and
$\Lambda_{\rm LP} = 100\,$TeV (orange cells). The red cells correspond
to representations that have a Landau pole below $100\,$TeV. In this work, we will consider that models with no Landau pole below $10^{10}$ GeV are acceptable, which leaves some room for other, heavier degrees of freedom to address the Landau pole problem.

\subsection{Lagrangian}
\label{sec:lagrangian}

The generic form of the Lagrangian we consider is
\begin{equation}
\cL \supset -m_D \psi\tilde{\psi} -{1\over 2} m_M \chi\chi  - y_1\psi \chi H^*  - y_2\tilde{\psi}\chi H + \mbox{\rm h.c.}
\label{eq:LDM}
\end{equation}
together with the kinetic terms of the new degrees of freedom. 
We take the Yukawa couplings to be real. We use the $SU(2)$
tensor formalism so that appropriate contractions of indices are
assumed. It may be useful to explicitly discuss a few examples. Writing the
components of the Higgs doublet as $H = (\phi^+,\phi^0)^T$, the
simplest case is the Yukawa coupling of two Weyl doublets, $\psi_i$
and $\tilde \psi_i$ with $i=1,2$, and one Majorana singlet $\chi$ or Bino-Higgsino system, to which we will refer as $1_M 2_D$, \begin{eqnarray} && -y_1\chi\psi_i
H ^{*i}=-y_1(\phi^{0*}\chi^0\psi^0+\phi^{+*}\chi^0\psi^+)\,,\nn
\\ &&-y_2\chi\tilde{\psi}_i
H_j\epsilon^{ij}=-y_2(\phi^{0}\chi^0\tilde{\psi}^0-\phi^{+}\chi^0\tilde{\psi}^-)\,.
\eea The next instance is the 
doublet-triplet system ({\em i.e.} Wino-Higgsino) or $3_M2_D$. The Weyl fermions are as
above, while the Majorana triplet is represented by an $SU(2)_L$ symmetric
tensor with 2 indices, $\chi_{ij} = \chi_{ji}$. The correspondence
between the tensor basis and the more
familiar basis in terms of eigenmodes of the $T_3$ generators ($T_3$
basis below) is easy to work out. For the $3_M$ we have
 \begin{equation}
\label{eq:3M}
\left(\!\begin{array}{c}\chi_{11} \\\sqrt{2}\chi_{12}
\\\chi_{22}\end{array}\!\!\!\right)\equiv\left(\!\begin{array}{c}\chi^{+}
\\\chi^{0} \\\chi^{-} \end{array}\!\right)\,,   \end{equation} and the Yukawa
couplings then take the form \bea && -y_1\psi_i\chi_{i'j}
H^{*j}\epsilon^{i'i}=-y_1({1\over
  \sqrt{2}}\phi^{0*}\chi^0\psi^0-\phi^{0*}\chi^-\psi^++\phi^{+*}\chi^+\psi^0-{1\over
  \sqrt{2}}\phi^{+*}\chi^0\psi^+)\,,\nn
\\ &&-y_2\tilde{\psi}_i\chi_{i'j}
H_{j'}\epsilon^{ii'}\epsilon^{jj'}=-y_2({1\over
  \sqrt{2}}\phi^{+}\chi^0\tilde{\psi}^--\phi^{+}\chi^-\tilde{\psi}^0+{1\over
  \sqrt{2}}\phi^{0}\chi^0\tilde{\psi}^0-\phi^{0}\chi^+\tilde{\psi}^-)\nn\,.
\end{eqnarray} The other cases are compiled in
Appendix~\ref{sec:tensor-formalism}.


The above combination of bare masses and Yukawa couplings gives rise
to mass matrices $M_Q$ for a set of fermions of charge $Q = T_3 + Y$
that take the same form for all the models studied here and are
uniquely determined by group representation. In e.g. the basis
$\{\chi^Q,\psi^Q,\tilde \psi^Q\}$, in the cases where 3 fermions
appear to have the same charge $Q$, $M_Q$ is given by:
 \begin{equation}
 M_Q^{3\times
   3}\!=\!(-1)^Q\left(\!\!\begin{array}{ccc} m_M & a_Q\hat{m}_1 &
 \tilde{a}_Q\hat{m}_2 \\ \tilde{a}_Q\hat{m}_1 & 0 & m_D \\ a_Q
 \hat{m}_2 & m_D & 0
 \end{array}\!\!\right)\, ,
 \label{eq:massmatrices}
 \end{equation}
while for one or two states of charge $Q$, $M_Q$ take the form: 
\begin{equation}
 M_Q^{2\times 2}\!=\!(-1)^Q\left(\!\!\begin{array}{cc}
 m_M & \hat{m}_1 \\
  \hat{m}_2 & m_D  \\
 \end{array}\!\!\right)\,, \quad
 M_Q^{1\times 1}\!=\!(-1)^Qm_D\,,
\label{eq:massmatrices2x2}
 \end{equation}
with $\hat{m}_{1,2}=y_{1,2}v/\sqrt{2}$ (with $v=246\,$ GeV) and
\begin{equation}
\label{eq:aQ}
a_Q=\textrm{min}\Big[{n_{\chi^{(-Q)}}\over n_{\psi^{Q}}},{n_{\psi^{Q}}\over n_{\chi^{(-Q)}}}\Big]\,,\quad \tilde{a}_Q=\textrm{min}\Big[{n_{\chi^{(-Q)}}\over n_{\tilde{\psi}^{Q}}},{n_{\tilde{\psi}^{Q}}\over n_{\chi^{(-Q)}}}\Big]\,.
\end{equation}
Here the $n_{\Psi^{(\pm Q)}}$ is the normalization factor that relates
a given component of a multiplet $\Psi$ of charge $\pm Q$ in the
tensor basis to that in the $T_3$ basis, as given in
Appendix~\ref{sec:tensor-formalism}. For instance, from (\ref{eq:3M})
we have for the triplet $\sqrt{2} \chi_{12} \equiv \chi^0$ and thus $n_{\chi^0} =
\sqrt{2}$, while $\chi_{22} \equiv \chi^+$ and so
$n_{\chi^+}=1$. For Yukawa couplings between a triplet and doublets,  $a_0  = \tilde a_0 = 1/\sqrt{2}$.

\subsection{Mass spectra}
\label{sec:mass-spectra}

To discuss the mass spectra we will exploit the existence of a global
$SU(2)_R$ symmetry\footnote{We follow here the nomenclature of the SM, in which the global symmetry acts naturally on right-handed fermions, {\em i.e.} SM $SU(2)_L$ singlet fermions.}, that mixes $\psi$ and $\tilde \psi$ when $y_1 = \pm
y_2$, to which we will refer as custodial points (see {\em e.g.}  \cite{Tait:2016qbg}).  A practical interest of that symmetry is that one can have rather transparent and simple analytic expressions for the mass spectrum and mixing matrices (at
least at tree level). More physically, we will see that it implies
that, after EW symmetry breaking, the particles fall into multiplets
of the diagonal subgroup {$SU(2) \subset SU(2)_L \times
  SU(2)_R$}. Away from $y_1 = \pm y_2$,
 the mass eigenstates are split but, thanks to the custodial symmetry, we will see that they remain nearly degenerate and thus can still be classified in terms of
{$SU(2)$} multiplets. 

We begin by considering the custodial limit, and then discuss in qualitative terms the more general situation. In principle we only need to  consider the case $y_1=y_2$ as, through the field redefinition $\tilde \psi \rightarrow - \tilde \psi$,  $y_1 = - y_2$  is equivalent to $y_1 = y_2$ together with a flip in sign of the Dirac mass, $m_D \rightarrow - m_D$.  However, we find it more convenient to fix the sign of $m_D$ and let the Yukawa couplings to have arbitrary signs. 

\subsubsection{Neutral states}
\label{sec:neutral-states}

Setting $y_1 = y_2 = y$ the mass matrix of neutral states is  diagonalized by going from the basis
$\xi_i=\{\chi_0,\psi_0,\tilde \psi_0\}$ with ${\cal L}_m=-\frac12
\sum_{ij}M_{0,ij} \xi_i\xi_j$ to the basis
$\chi_i=\{\chi^0_1,\chi^0_2,\chi^0_3\}$ with ${\cal L}_m=-\frac12
\sum_{i}m_{i} \chi_i\chi_i$ and
\begin{eqnarray}
  m_1&=&\frac12 (m_M+m_D+\Delta m_\eta)\nonumber\\
  m_2&=&m_D\\
  m_3&=&\frac12(m_M+m_D-\Delta m_\eta)\nonumber
\label{eq:mcust}
\end{eqnarray}
where
\begin{equation}
  \Delta m_\eta=\sqrt{(m_D-m_M)^2+ 8 (\eta y v/\sqrt{2})^2}\,.
\label{eq:dmcust}
\end{equation}
Notice that $\eta$ is  equal to the coefficients $a_{Q=0}=a_{\tilde Q=0}$
that appear in the $M_0$ mass matrix of
eq.~(\ref{eq:massmatrices}). In particular, for the cases that we are interested in,
we have:
\begin{equation}
\eta= 
\left\{
\begin{array}{lllc}
1 & & &1_M 2_D\\
 {1/\sqrt{2}}& & &3_M 2_D\;\&\; 5_M 4_D\\
\sqrt{2/3} & & &3_M 4_D
\end{array} \right.  
\end{equation}
  For the diagonalisation, we use the
 transformation \footnote{This transformation matrix comes from the fact that for $y_1 = y_2$ only the combination $\chi' \sim \psi +\tilde \psi$
    couples to the Higgs. One obtains
     (\ref{eq:rot}) combining a $\pi/4$ rotation of the states
    $\psi$ and $\tilde \psi$ together with a rotation of angle
    $\theta_\eta$  in the subspace spanned by  $\chi'$ and $\chi$. \label{foot:rotation}}
\begin{equation}
  \left(\!\begin{array}{c} \chi_1^0\\\chi_2^0  \\ \chi_3^0\end{array}\!\right)
  \!=\!
  \left(\!\!\begin{array}{ccc}
 c_\eta& s_\eta/\sqrt{2} &s_\eta/\sqrt{2}\\
 0&i/\sqrt{2}&-i/\sqrt{2}\\
 -s_\eta& c_\eta/\sqrt{2} &c_\eta/\sqrt{2}\\
  \end{array}\!\!\right)\!
  \left(\!\begin{array}{c} \chi_0\\\psi_0  \\ \tilde \psi_0\end{array}\!\right)
 \label{eq:rot}
\end{equation}
with $s_\eta= \sin \theta_\eta$ and $c_\eta= \cos\theta_\eta$ considering
\begin{equation}
 \sin^2\theta_\eta=\frac12 \left( 1+ \frac{m_D-m_M}{\Delta m_\eta}\right)\,.
 \label{eq:mixingangle}
\end{equation}
The transformation matrix used in eq.~(\ref{eq:rot}) is equivalent to
the one of~\cite{Freitas:2015hsa} up to some differences in
normalization and sign conventions. In addition, our $\chi^0_i$
indices $i=1,2,3$ do not point to any mass ordering. The latter
depends on the hierarchies between $m_D$ and $m_M$ and between $\eta
yv $ and $\sqrt{m_D^2-m_Mm_D}$. Going from the basis above to the mass
ordered basis $\{\chi^0_\alpha\}$ with indices $\alpha=l,m,h$
(refering to the light, medium and heavy states) just simply imply a
reordering of the transformation matrix entries. The Lagrangian with
couplings to the Higgs ($h$) and the Z boson takes the form
\begin{eqnarray}
  \cal L&=& -\frac{g}{2} (\psi_0^\dag\bar\sigma^\mu \psi_0-\tilde \psi_0^\dag\bar\sigma^\mu \tilde \psi_0) Z_\mu-y\eta (\tilde \psi_0-\psi_0) \chi_0 \,h ,
  \label{eq:lint}
\end{eqnarray}
which corresponds in the basis of mass eigenstates to
\begin{eqnarray}
  \cal L&=& \frac{g}{2} \chi_2^{0*} \bar\sigma^\mu (s_\eta \chi_1^{0}+ c_\eta \chi_3^{0})  Z_\mu+h.c.\cr
  &&-\frac{y\eta}{2\sqrt{2}} \left (s_{2\eta}(  \chi_1^0\chi_1^0-\chi_3^0\chi_3^0)+2 c_{2 \eta}\chi_1^0\chi_3^0\right)\,h +h.c.
  \label{eq:lmassb}
\end{eqnarray}
with $s_{2\eta}=\sin(2\theta_\eta)$ and
$c_{2\eta}=\cos(2\theta_\eta)$. This is in agreement with
\cite{Freitas:2015hsa} for $1_M2_D$ and $3_M 2_D$, up to distinct phase conventions.\footnote{Notice that we do not obtain a $1/c_w$ prefactor in the $Z_\mu$ coefficient.} 
  
  We first briefly comment on the
above Lagrangian, as it will be of interest for DM scattering on
nucleons. First of all, the couplings to the $Z$ are non-diagonal reflecting the fact that, unless $y=0$, the mass eigenstates are all Majorana particles. The constraints from direct dark matter searches are thus avoided provided the mass differences between $\chi_{1,3}$ and $\chi_2$ are larger than ${\cal O}$(100 keV) \cite{TuckerSmith:2001hy}. Notice also that  one of the neutral particles (here $\chi_2$) does not couple to the Higgs. This feature is also generic, as only the combination $\sim y_1 \psi + y_2 \tilde \psi$  mixes with the Majorana multiplet (see also footnote \ref{foot:rotation}).  Then there are some potentially interesting limiting cases (see also~\cite{Freitas:2015hsa}): 
\begin{itemize}
\item From (\ref{eq:mcust}) and (\ref{eq:mixingangle}) we see that the
  Lightest Neutral Particle (LNP) has maximal coupling to the Higgs
  when \underline{$m_M\simeq m_D$ and $y_1 \simeq y_2$} with
  $y_1,y_2\gg |m_M-m_D|/(2\sqrt{2}\eta v)$.  Indeed at the custodial
  point $y_1= y_2$ and $m_N=m_D$ so that $\chi_3^0 = \chi_0$ is the DM
  candidate and $\theta_\eta= \pi/4$.  Moving away from this custodial
  point, we have checked numerically that the coupling to the Higgs
  remains close to maximal coupling when $m_M\simeq m_D$, $y_1$ and
  $y_2$ have the same sign and $\vert y_1 + y_2\vert \gg 1$.
    
  \item In the limit \underline{ $m_D\gg m_M$ and $y_1 \simeq y_2$}  with $y_1,y_2\ll  |m_M-m_D|/(2\sqrt{2} \eta v)$ 
  one recovers the case of the
    Majorana DM case with zero coupling to the Higgs and kinematically
    suppressed coupling to the $Z$. Indeed, at the custodial point $y_1 = y_2$ ($y_1 = -y_2$),  $\chi_3^0 = \chi_0$ (resp. $\chi_1^0 = \chi_0$) is the DM candidate and $\theta_\eta= \pi/2$ (resp. $\theta_\eta = 0$). 
  \item When \underline{ $m_M\gg m_D$ and $y_1 \simeq y_2$} with small enough Yukawa couplings the states
    $\chi_0^3$ and  $\chi_0^2$ have a mass splitting  $\delta m = {\cal O}(y^2 v^2/m_N)$, forming a pseudo-Dirac fermion and their
    coupling to the $Z$ is maximal as $\theta_\eta \simeq 0$. As usual, to avoid constraints from direct detection, the mass splitting must satisfy $\delta m > {1/2} \mu v^2 \sim 100$ keV, where $v \simeq 10^{-3}$ is the velocity of the dark matter and $\mu$ is the reduced mass of the dark matter/direct detection target nucleus \cite{TuckerSmith:2001hy}, see Sec.~\ref{sec:direct-detect-search} for more details. 
  \item Finally, let us stress that for \underline{ $m_M\simeq m_D$
    but $y_1\simeq -y_2$}, i.e. with Yukawas of opposite signs, the
    lightest neutral state has suppressed coupling to the Higgs. This
    can be seen from Eqs. (\ref{eq:mcust}) and (\ref{eq:mixingangle}),
    obtained in the limit $y_1=y_2=y$, by setting $m_D \rightarrow -
    m_D$. In the latter case,  the LNP is
    $\chi_2^0$ and corresponds to the combination of Weyl states { $\propto \psi_0-\tilde
    \psi_0$ } that does not couple to the Higgs. As one departs from
    this custodial point, the LNP mixes with the neutral component of the
    Majorana multiplet, $\chi_0$,  and so couples to
    the Higgs.\footnote{To leading order in $y_1 + y_2$ the
      mass of the LNP does not change but the mixing with $\chi_0$ is
      $\propto \delta m/m \times m_M/m$ where $m \simeq y v$ and
      $\delta m \sim (y_1 + y_2) v$.}  We have checked numerically
    that this behavior holds over a broad range of parameters away
    from the custodial point $y_1 = - y_2$.
  
\end{itemize}

\subsubsection{Charged states and $SU(2)$ multiplets structure}
\label{sec:charg-stat-mult}

We now comment on the mass spectrum of the charged partners. As
mentioned above, at the custodial points the neutral, singly charged
and, if present, doubly charged eigenstates combine into multiplets of
the custodial {$SU(2)$}. Of course, the custodial symmetry is only
approximate, being explicitly broken by coupling to $U(1)_Y$ gauge
bosons. In the case of Minimal Dark Matter, one-loop electroweak
corrections induce splittings ${\cal O}$(100 MeV) between the
components of a multiplet such that the neutral state of a multiplet
with $Y=0$ is always the lightest component, and so is potentially a
dark matter candidate~\cite{Cirelli:2005uq}, see
also~\cite{McKay:2017rjs} for a recent discussion.  Once non-zero
Yukawa couplings between different representations are considered,
there are more possibilities as, away from the custodial points, mass
splittings between components are obtained already at tree level. We
first focus on tree-level splittings and then comment on the potential
effects of loop corrections.

A first feature is that for $y_1 = \pm y_2$, the Majorana and two Weyl states mix and, together, neutral and charged particles combine to form Majorana {$SU(2)$} multiplets according to the following pattern:
\begin{eqnarray}
1_M 2_W 2_W \rightarrow  1_M 1_M 3_M &\qquad & 3_M2_W 2_W\rightarrow 3_M 1_M 3_M \nonumber \\
\\
  3_M 4_W 4_W\rightarrow 3_M 3_M 5_M&\qquad & 5_M 4_W 4_W \rightarrow 5_M 3_M 5_M \, .\nonumber
  \label{eq:pattern}
\end{eqnarray}
In essence, the two n-plet Weyl states (of same chirality and thus opposite hypercharge) combine to form a Majorana $(n+1)$-plet, the orthogonal state being a Majorana $(n-1)$-plet.  At the custodial points, the components of each multiplet are degenerate, but distinct multiplets have a distinct mass. The multiplet that contains the dark matter candidate can be determined by direct evaluation of the mass eigenstates. However, as the mixing between three neutral states involves solving a cubic equation, the outcome is not a priori obvious. Fortunately, the mass spectra have some general features, which are easy to grasp using the custodial symmetry.

In what follows, we provide a detailed case by case study. In essence, the
relevant points of the discussion below can be summarized as follows:
1) at the custodial points, the LNP belongs in general (it can be in a
$1_M$, for instance in the $1_M 2_D$) to a multiplet of the $SU(2)$
custodial symmetry; 2) away from the custodial points, the multiplet
components are split, but the splitting is somewhat protected by the
custodial symmetry and 3) the LNP is always the lightest component of
the multiplet; 4) the mass splitting are ${\cal O}((y_1 \pm y_2)^2
v^2/m_M)$ if $m_M \gg m_D$ and ${\cal O}((y_1^2 - y_2^2) v^4/m_D^3)$
if $m_M \ll m_D$, assuming small Yukawa couplings.

\begin{table}[t]
    \begin{tabular}{ | c | c | c |}
    \hline
    M-D system & $m_M < m_\ast\sim m_D$ & $m_M > m_\ast \sim m_D$ \\ \hline\hline
   $1_M2_D \sim 1_M 1_M 3_M$ & $\chi_l^0 \sim  1_M$& $\chi_l^0 \subset \left\{\begin{array}{lll}
   3_M  & {\rm at} & y_1= - y_2 \\
   1_M & {\rm at} &y_1 =  y_2 \end{array}\right.$
    \\ \hline
    $3_M 2_D \sim 1_M 3_M 3_M$ & $\chi_l^0 \subset  3_M$ & $\begin{array}{lll} 
   1_M  & & y_1= - y_2 \\
   3_M && y_1 =  y_2 \end{array}$
     \\ \hline
    $3_M 4_D \sim 3_M 3_M 5_M$ & $3_M$ & $\begin{array}{lll} 
   5_M  & &y_1= - y_2 \\
   3_M && y_1 =  y_2 \end{array}$ \\
    \hline
    $5_M 4_D \sim 3_M 5_M 5_M$ & $5_M$ & $\begin{array}{lll} 
   3_M  & &y_1= - y_2 \\
   5_M & &y_1 =  y_2 \end{array}$\\
   \hline
    \end{tabular}
    \caption{After EWSB the Weyl and Majorana states mix. At the custodial points ($y_1 = \pm y_2$) they combine into multiplets of a custodial $SU(2)$ symmetry. Away from the custodial points, the multiplets component are split, but remain nearly degenerate, thanks to the custodial symmetry. See text. The table shows to which $SU(2)$ multiplet the LNP (lightest neutral particle)  $\chi_l^0$ belongs for each case. This depends on the mass hierarchy between the bare Majorana and Dirac masses, or more precisely on whether $m_M$ is smaller or larger than  $m_\ast   =  m_D - y_1^2 (\eta v)^2/2 m_D$.     }
  \label{tab:charged}
\end{table}

\paragraph{The $1_M2_D$ case}

\begin{figure}[t!]
  \begin{center}     
  \begin{tabular}{cc}
      \includegraphics[width=7.5cm]{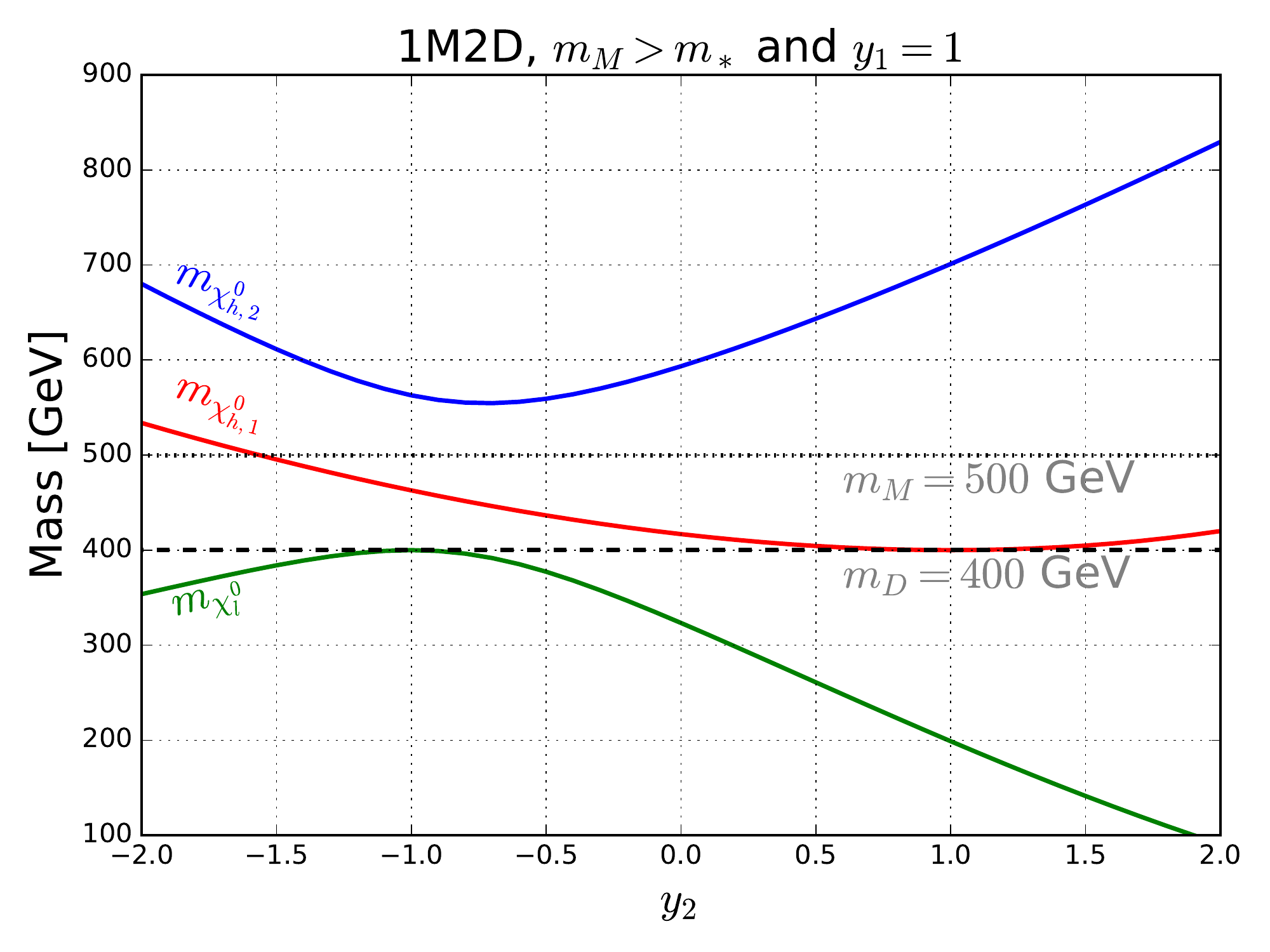}
       \includegraphics[width=7.5cm]{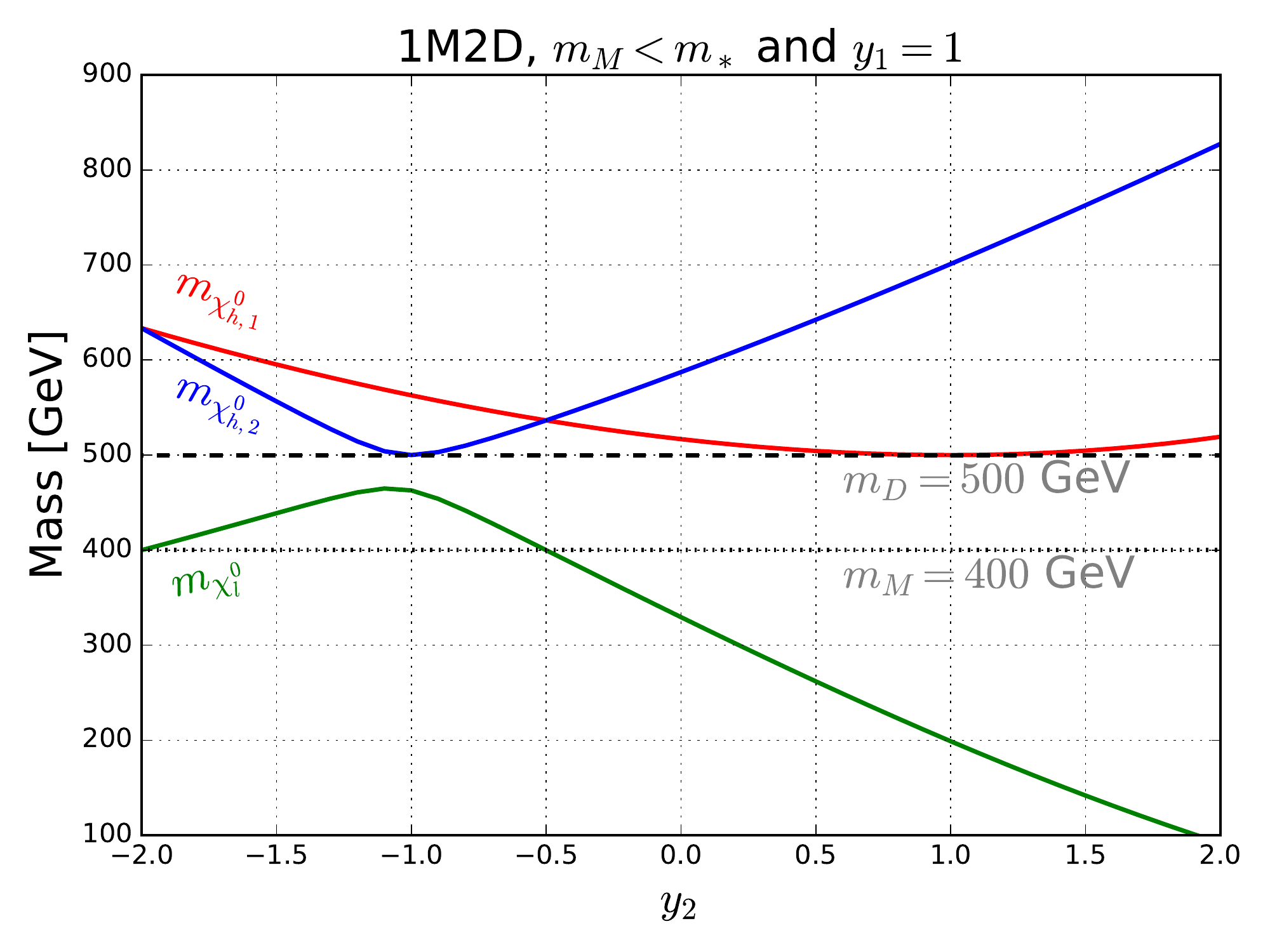}
       \end{tabular}
  \end{center}
  \caption{Mass spectra on the $1_M 2_D$ system for $y_1 = 1$ as a function of $y_2$. The masses of the neutral states are depicted with continuous colored lines and by a black dashed line for the charged components. We use the subscripts $l$ and $h_{1,2}$ to refer respectively to light and heavy neutral eigenstates.  These spectra illustrate the fact that the charged states combined with a singlet to form a Majorana triplet  $3_M$ at the custodial points $y_1 = \pm y_2$.  The lightest neutral particle (LNP $\sim \chi_l^0$) is in this case generically a Majorana singlet, except near the custodial point $y_2 = - y_1\equiv - 1$ if $m_M > m_\ast \sim m_D$ where it forms a $3_M$. There is another $3_M$ at $y_2 = y_1 \equiv 1$ but its neutral component is not the LNP.  See text for more details. }
\label{fig:1M2Dtree}
\end{figure}

\begin{figure}[t!]
  \begin{center}     
  \begin{tabular}{cc}
      \includegraphics[width=7.5cm]{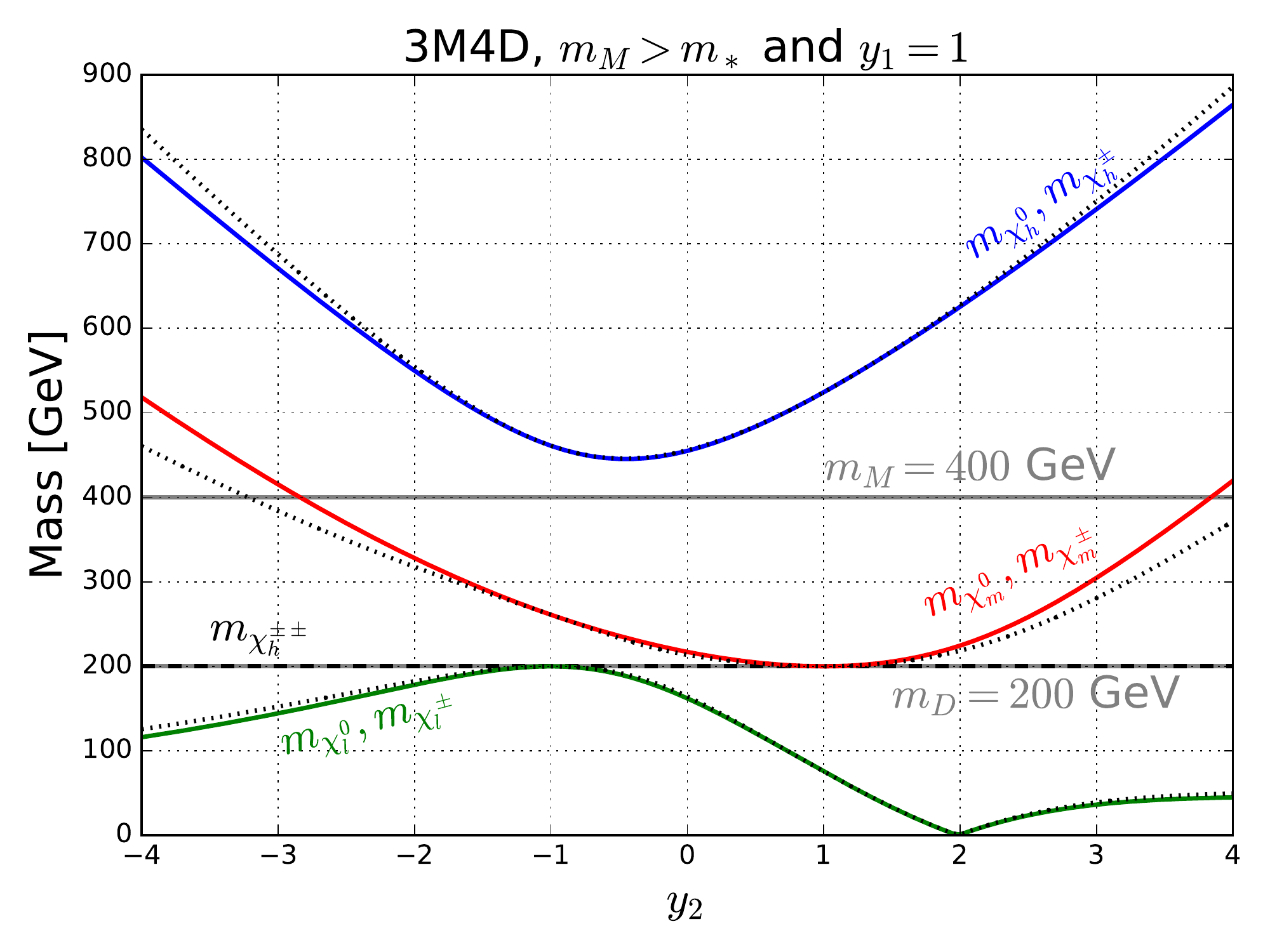}
       \includegraphics[width=7.5cm]{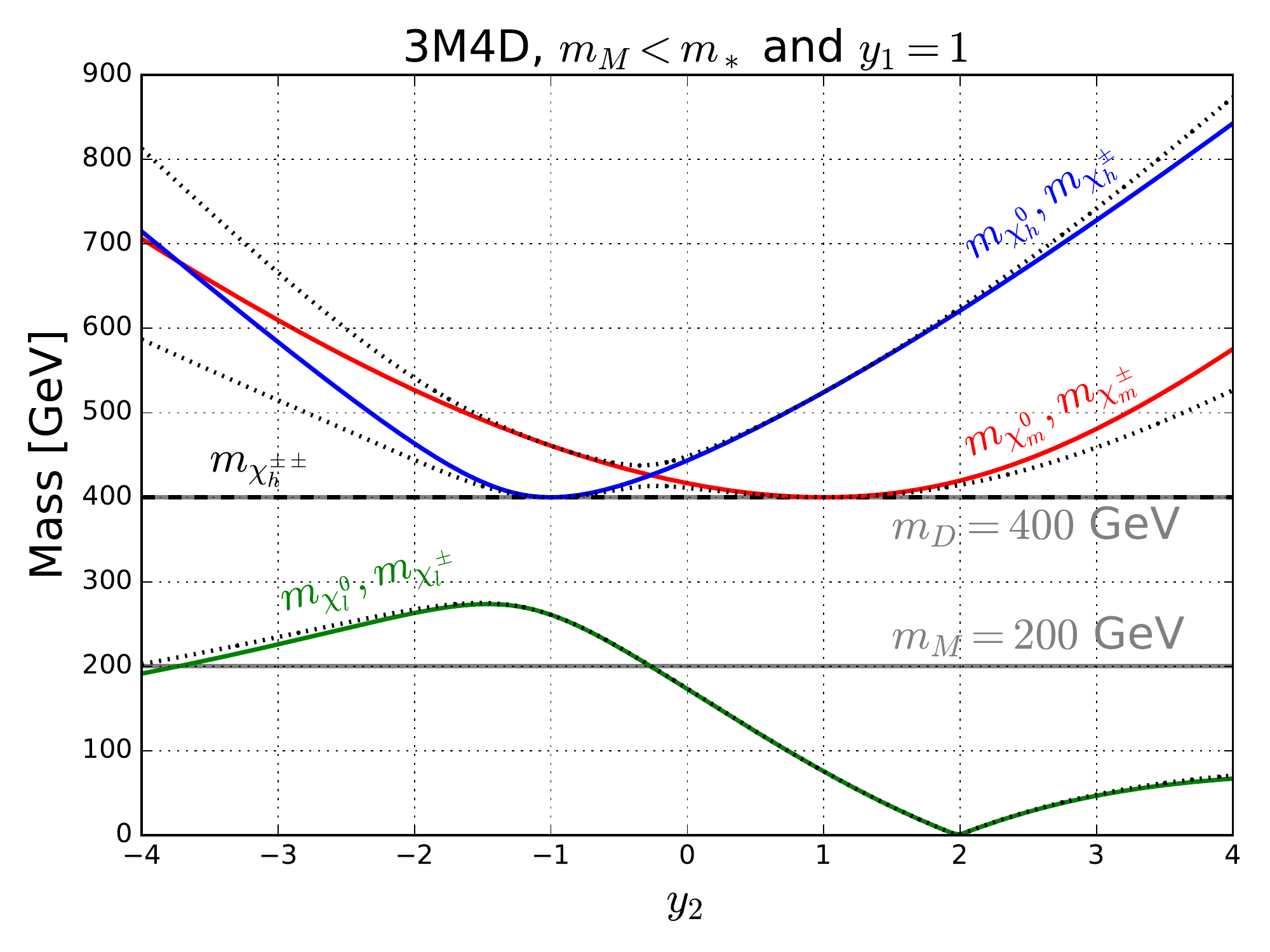}
       \end{tabular}
  \end{center}
  \caption{Mass spectra on the $3_M 4_D$ system for $y_1 = 1$ as a
    function of $y_2$. The eigenmass of the neutral states are
    depicted with the same color coding as in Fig.\ref{fig:1M2Dtree}; singly charged eigenmass are shown as black dotted lines; the doubly charged states have constant mass $m_D$. These spectra are meant to illustrate the
      fact that the neutral and charged states combine in 2
      Majorana  $3_M$ and one $5_M$ of $SU(2)$ at the
      custodial points $y_2 = \pm y_1 \equiv \pm 1$ and, also, that they are nearly degenerate away from these points. }
\label{fig:3M4D}
\end{figure}

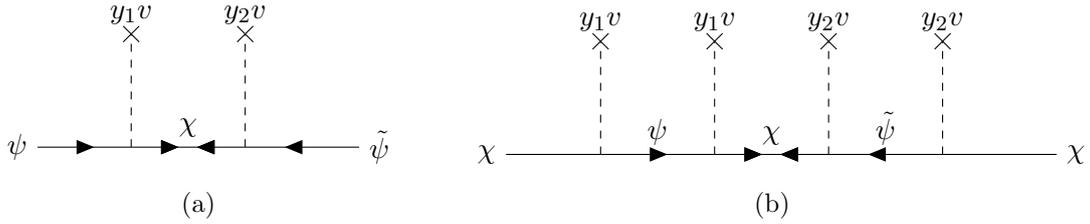
\begin{figure}[t!]
\centering
\begin{subfigure}[b]{0.35\textwidth}
\centering
\begin{tikzpicture}
\begin{feynman}
\vertex (a) {\(\psi\)};
\vertex [right=of a] (b);
\vertex [above=of b] (s1) {\( y_1v\)};
\vertex [right=of b] (c) ;
\vertex [above=of c] (s2) {\( y_2 v\)};
\vertex [right=of c] (f3) {\(\tilde \psi\)};
\diagram* {
(a) -- [fermion] (b) -- [scalar,insertion=1] (s1),
(b) -- [majorana, edge label=\(\chi\)] (c),
(c) -- [scalar,insertion=1] (s2),
(c) -- [anti fermion] (f3),
};
\end{feynman}
\end{tikzpicture}
\subcaption{}
\label{fig:SplitmMgmD}
\end{subfigure}
\qquad
\begin{subfigure}[b]{0.53\textwidth}
\centering
\begin{tikzpicture}
\begin{feynman}
\vertex (a) {\(\chi\)};
\vertex [right=of a] (b);
\vertex [above=of b] (s1) {\( y_1 v\)};
\vertex [right=of b] (c) ;
\vertex [above=of c] (s2) {\( y_1 v\)};
\vertex [right=of c] (d);
\vertex [above=of d] (s3) {\( y_2v\)};
\vertex [right=of d] (e);
\vertex [above=of e] (s4) {\( y_2 v\)};
\vertex [right=of e] (f){\(\chi\)};
\diagram* {
(a) -- [plain] (b) -- [scalar,insertion=1] (s1),
(b) -- [fermion, edge label=\(\psi\)] (c),
(c) -- [scalar,insertion=1] (s2),
(c) -- [majorana, edge label=\(\chi\)] (d),
(d) -- [scalar,insertion=1] (s3),
(d) --  [anti fermion, edge label=\(\tilde \psi\)] (e),
(e) -- [scalar,insertion=1](s4),
(e) -- [plain](f),
};
\end{feynman}
\end{tikzpicture}
\subcaption{}
\label{fig:SplitmMlmD}
\end{subfigure}
\caption{Contributions to mass splitting within n-plets for $m_M \gg m_D$ (\ref{fig:SplitmMgmD}) and $m_D \gg m_M$ (\ref{fig:SplitmMlmD}).}\label{fig:massSplits}
\end{figure}

\begin{figure}[t!]
  \begin{center}     
    \begin{tabular}{cc}
      \includegraphics[width=7.5cm]{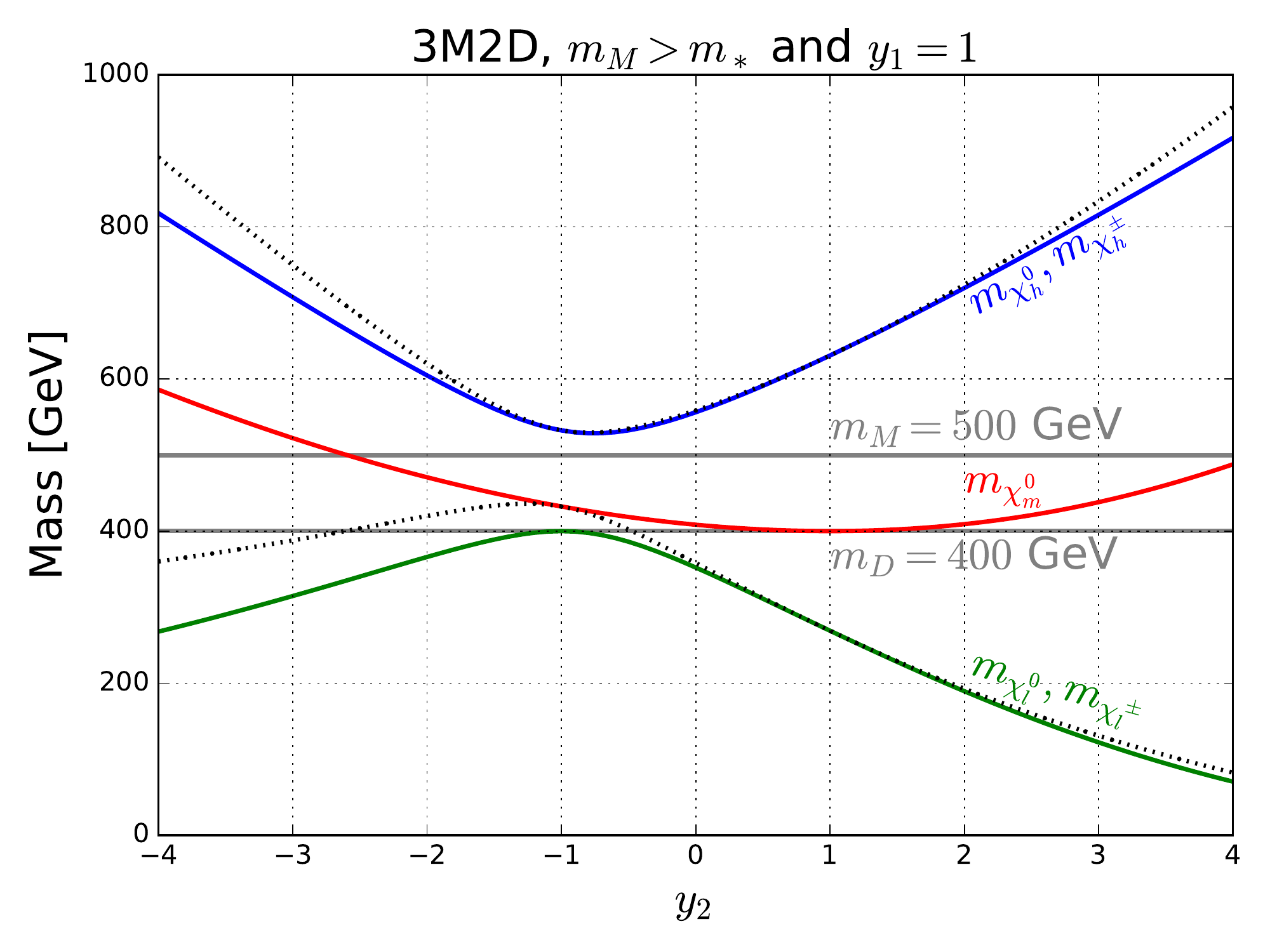}&
      \includegraphics[width=7.5cm]{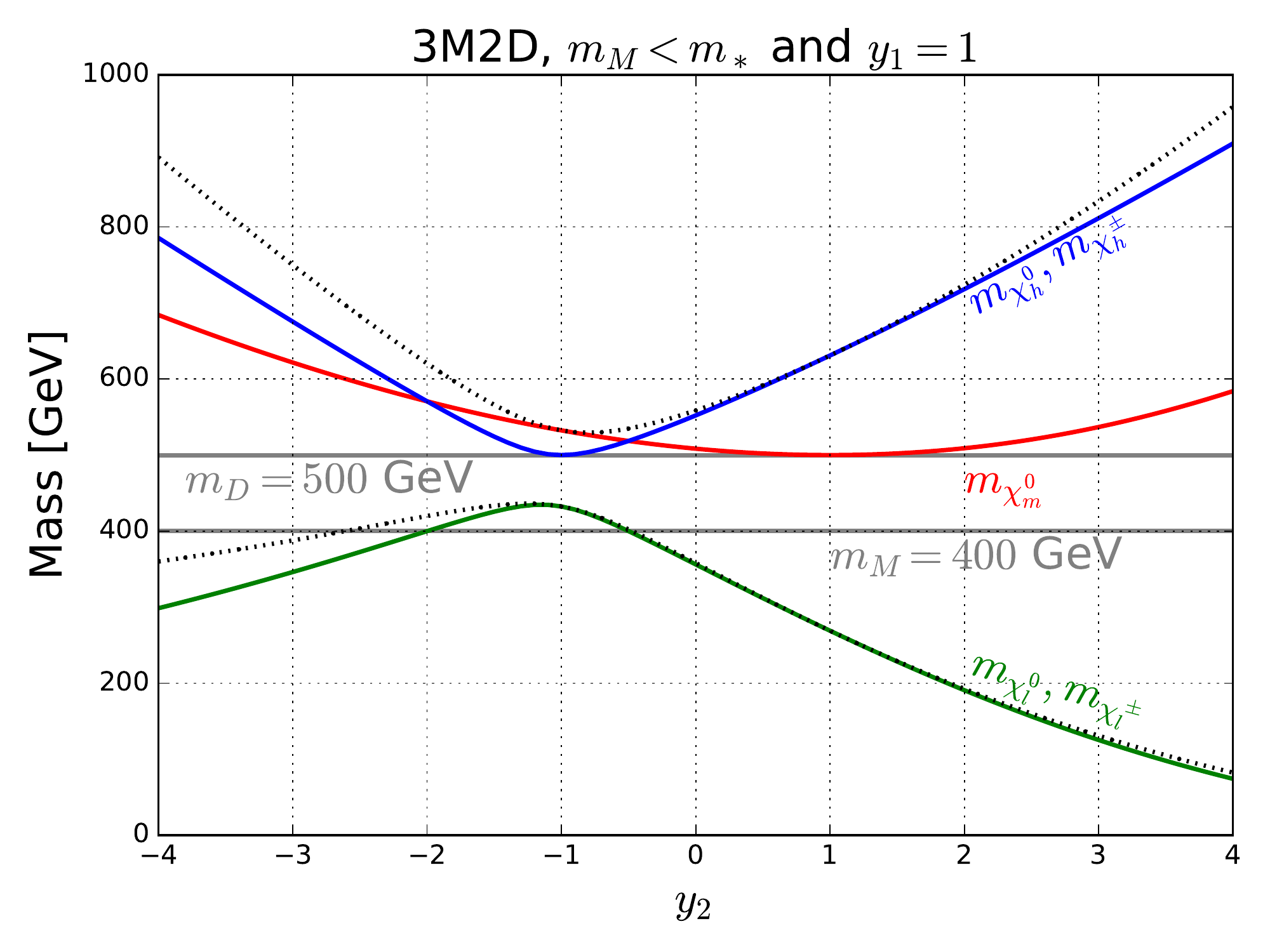}
    \end{tabular}
  \end{center}
  \caption{ Mass spectra  in the $3_M2_D$ system for
    $y_1=1$ as a function of $y_2$ for $m_M \gtrsim m_\ast$ (left panel) and $m_M \lesssim m_\ast$ (right panel). Masses of neutral states are
    depicted with continuous colored lines and  the singly charged
    states with black dotted lines. }
    \label{fig:3M2D}
\end{figure}

\begin{figure}[t!]
  \begin{center}     
    \begin{tabular}{cc}
      \includegraphics[width=7.5cm]{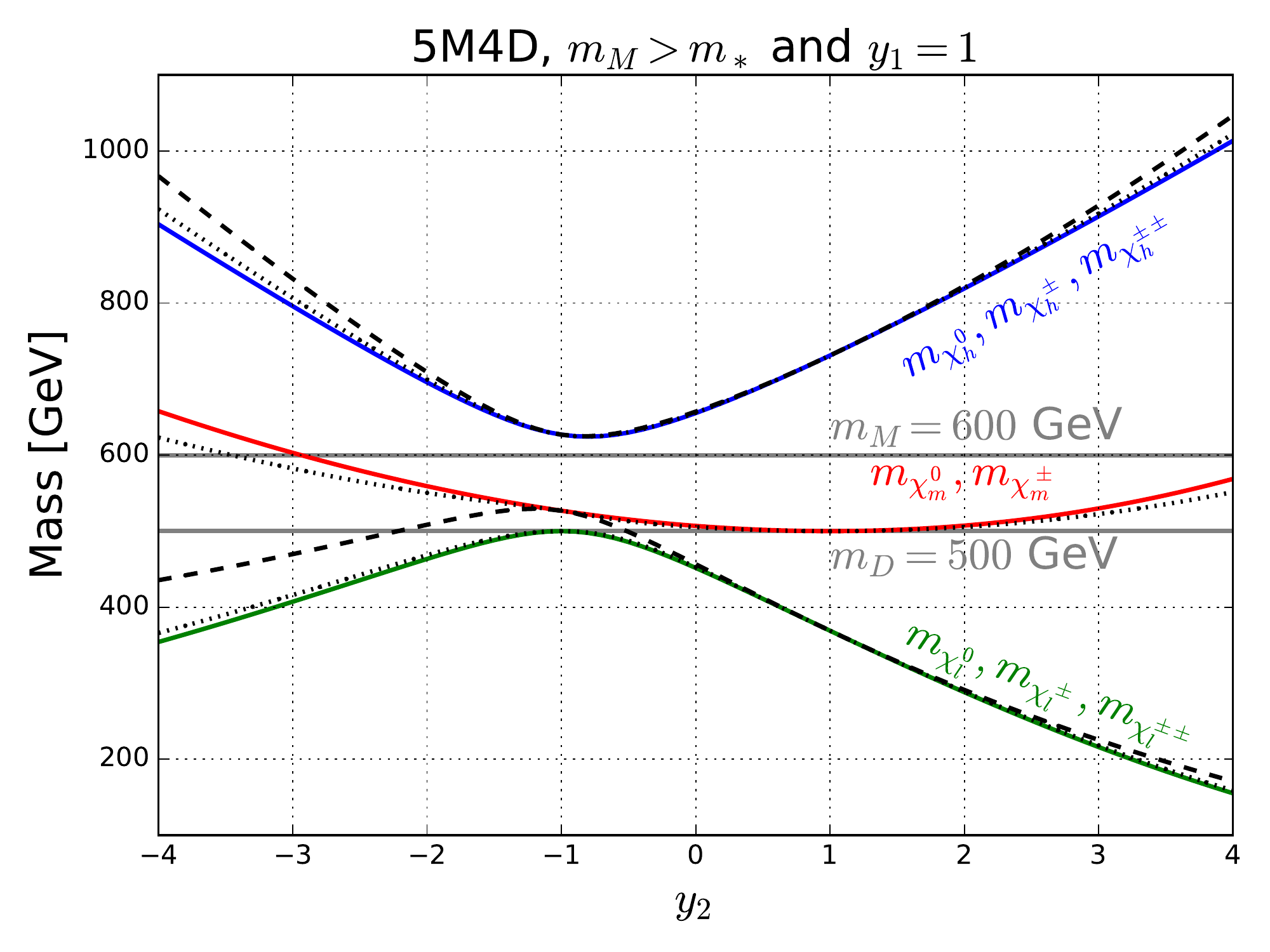}&
      \includegraphics[width=7.5cm]{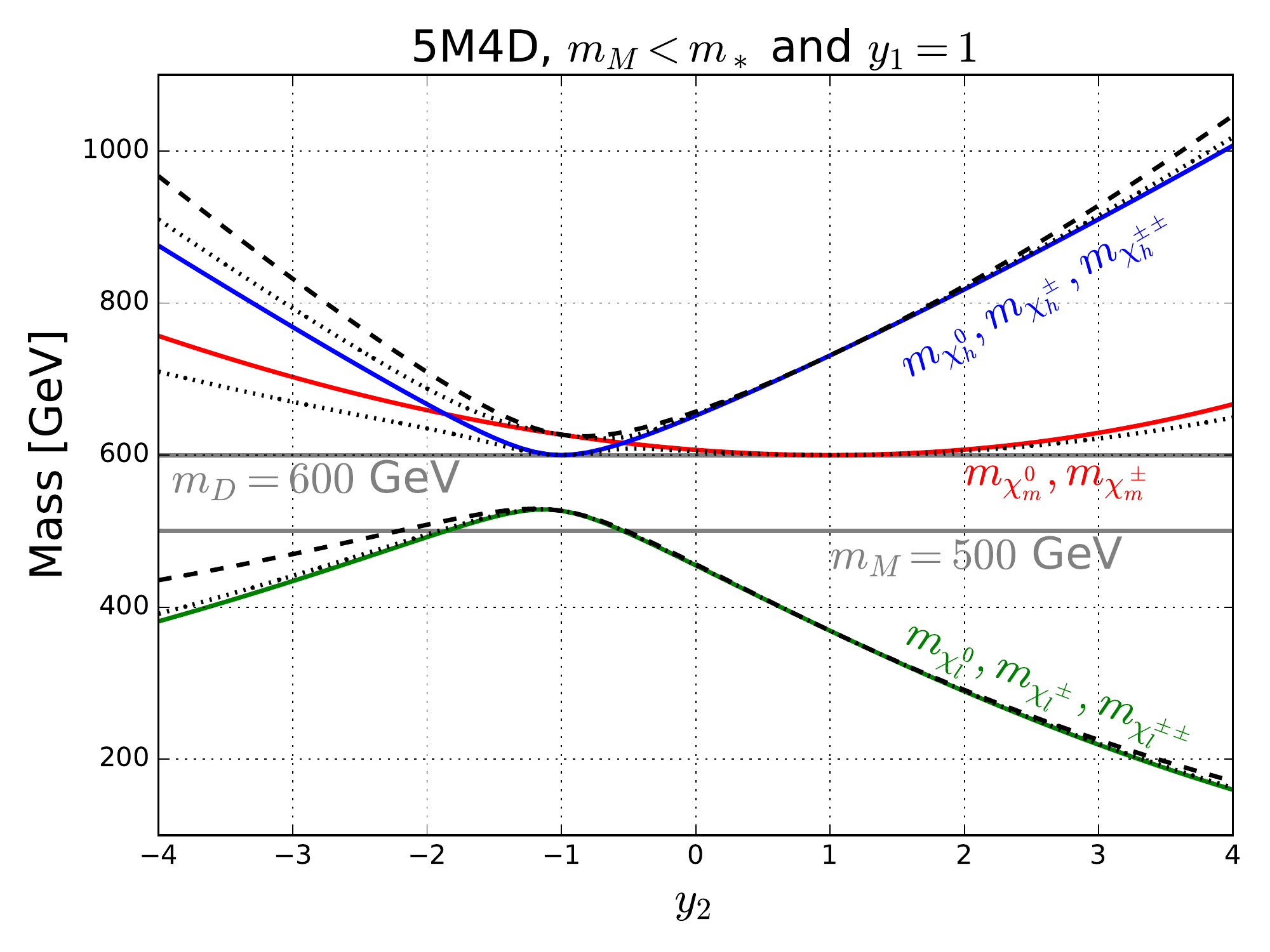}
    \end{tabular}
  \end{center}
  \caption{Mass spectra  in the $5_M4_D$ system for
    $y_1=1$ as a function of $y_2$ for $m_M \gtrsim m_\ast$ (left panel) and $m_M \lesssim m_\ast$ (right panel). Masses of neutral states are
    depicted with continuous colored lines, for the singly charged
    states with black dotted lines and for the doubly charged states
    with black dashed lines. }
    \label{fig:5M4D}
\end{figure}

 The typical spectra are shown in Fig.\ref{fig:1M2Dtree} for the cases $m_M \gtrsim m_D$ (left panel) and $m_M \lesssim m_D$ (right panel).\footnote{Note that in Fig.\ref{fig:1M2Dtree}, and especially in Figs.~\ref{fig:3M4D}-\ref{fig:split3M4D}, we use parameters that do not specifically refer to viable DM candidates but are meant to clearly  illustrate our discussion of the mass spectra.}  
In each panel, the three solid lines correspond to the three neutral
states, the lightest being a potential DM candidate. The horizontal
dashed line corresponds to the charged states, with $m_{\chi^{\pm}} =
m_D$. Focusing on the custodial point $y_2 = -y_1$, we observe that
clearly two of the neutral states, one of which has mass $m_D$ at $y_2 =
-y_1$, have an avoided level crossing.\footnote{For clarity, we plot the absolute value
  of all the masses. The third neutral state, corresponding to the red
  lines in Fig.\ref{fig:1M2Dtree}, state has a negative eigenvalue
  mass (in our basis). In general, there is level repulsion between all the states.} The latter corresponds to the combination of Weyl states that does not
couple to the Higgs.
This state is degenerate with the charged states, and altogether
they form an $SU(2)$ triplet, $3_M$. Whether the LNP belongs to this
triplet depends on the hierarchy between the bare Dirac and Majorana masses, $m_M \gtrsim m_D$ (left panel) or $m_M
\lesssim m_D$ (right panel). More precisely, it is easy to verify   that the levels cross
when
 \begin{equation}
 m_M = m_\ast \equiv m_D - y_1^2\,  {\eta^2 v^2\over 2 m_D}
\end{equation}
were we assumed $y_1 v \ll m_D$ with $y_2 = -y_1$.  
If $m_M > m_\ast$, the LNP has mass $m_D$ and, together with the charged states, is in a triplet, $3_M$. If instead $m_M < m_\ast$, the LNP is a singlet, $1_M$. The latter state is a mixture of the original Majorana singlet $\chi_0$ and of the combination of Weyl states  to which it couples through the Yukawa.  

Away from  the custodial point $y_2 = -y_1$, we observe from Fig.\ref{fig:1M2Dtree} that the mass eigenstates repel each other so that the mass of the LNP decreases while the mass of the charged partner stays constant, $m_{\chi^\pm} \equiv m_D$. Level repulsion thus explains why the LNP is also the lightest particle, and so potentially a dark matter candidate. For $m_M > m_\ast$ and working in the limit $\vert y_1 + y_2\vert v \ll m_{M,D}$, it is easy to obtain that the mass splitting is given by
$$
\Delta m = m_{\chi_l^\pm} -  m_{\chi^0_l}  \approx {a_0^2\over 4} (y_1 + y_2)^2 {v^2\over m_M}  \equiv (y_1 + y_2)^2{ v^2\over 4\, m_M} 
$$
where the subscript $l$ stands for "light". 
So for $y_1 + y_2 \neq 0$, the LNP is a singlet, and this both for $m_M > m_\ast$ and $m_M < m_\ast$.

Finally, from Fig.\ref{fig:1M2Dtree}  we notice that  the charged states combine with another singlet  at $y_1= y_2$. This triplet is however heavier than the LNP. \footnote{Also, we notice that the mass of this LNP may vanish for large enough Yukawa couplings. This happens if the  $m_M  m_D \approx \eta^2 y_1 y_2 v^2$ and so, assuming perturbative couplings, $y_{1,2} \lesssim 4 \pi$, only for $m_D m_M < {\cal O}$(TeV).} To recap, at the custodial points, the pattern of multiplet is as in (\ref{eq:pattern}), with $1_M 2_W 2_W \rightarrow  1_M 1_M 3_M$. Whether the LNP is in a $1_M$ or a $3_M$ is summarized in Table \ref{tab:charged}.

\paragraph{The $3_M4_D$ case}

We discuss this  next because it shares features with 
the $1_M 2_D$  case. According to (\ref{eq:pattern}), we have the pattern $3_M 4_W 4_W\rightarrow 3_M 3_M 5_M$ at the custodial points. This is illustrated in Fig. \ref{fig:3M4D} that shows that the neutral states follow always the same pattern as in the 
$1_M 2_D$ system discussed above. The question is what is the mass spectrum of the charged
partners? In the $3_M 4_D$ case, it is the doubly charged state $\chi^{\pm\pm}$ that does not mix and so has
  mass $m_D$. At the custodial point $y_1
= -y_2$ we observe from Fig.\ref{fig:3M4D} that it belongs to a $5_M$ formed with states
(neutral and singly charged) that do not couple to the Higgs. This
$5_M$ contains the LNP if $m_M > m_\ast$. If $m_M < m_\ast$, the
LNP is instead in a $3_M$. The twist compared to the $1_M 2_D$ case
is that, away from $y_1 = -y_2$, level repulsion brings down {\em
  both} the mass of the LNP and that of its singly charged partners,
so that the LNP belongs to a nearly degenerate $3_M$
multiplet. The reason for this interesting behavior may be understood analytically by considering the
hierarchies $y_{1,2} v \ll m_D \ll m_M$ or $y_{1,2} v \ll m_M \ll m_D$.

\begin{enumerate}
\item \underline{ $m_M \gg m_D$} At $y_1 = -
  y_2$,  the LNP belongs to a $5_M$ of $SU(2)$ with mass $m_D$. The
  doubly charged components do not mix, so the mass is equal to $m_D$ for all $y_{1,2}$. 
Away from the custodial point  $y_1 = -
  y_2$, level repulsion
  brings down the mass of both the neutral and singly charged
  components. In the limit $y_{1,2} v \ll m_D \ll m_M$, we get near $y_2 = -
  y_1$ that
\begin{eqnarray}
\label{eq:split1}
m_{\chi^0_l} &\approx& m_D - {a_0^2\over 4} (y_1 + y_2)^2 {v^2\over m_M}\nonumber\\
& \approx & m_D - {1\over 6} (y_1 + y_2)^2 {v^2\over m_M}
\end{eqnarray}
while
\begin{eqnarray}
m_{\chi^\pm_l} &\approx& m_D - {a_1^2 \tilde a_1^2\over  2(a_1^2 + \tilde a_1^2)} (y_1 + y_2)^2 {v^2\over m_M}\nonumber\\
& \approx & m_D - {1\over  8} (y_1 + y_2)^2 {v^2\over m_M}
\end{eqnarray}
Thus, the mass splitting between the singly charged components and the LNP is 
\begin{equation}
\label{eq:split2}
\Delta m = m_{\chi^\pm_l} - m_{\chi^0_l}\approx {1\over 24} (y_1 + y_2)^2 {v^2\over m_M} > 0
\end{equation}
and the LNP is, at tree level, the lightest component of a nearly degenerate $3_M$ {\em away from the custodial point}. This is a generic conclusion: in all cases, the LNP is at tree level always the lightest component of the $SU(2)$ multiplet to which it belongs, and thus {\em a priori} a DM candidate. Why this is so is a bit mysterious but may be traced to the entries in the mass matrices, see (\ref{eq:massmatrices}-\ref{eq:aQ}). The outcome is that, somehow, level repulsion is stronger for the neutral particles than it is for their charged partners. 
We also infer that the custodial symmetry is keeping the $3_M$ nearly degenerate. We interpret this as being due to the fact that at  the other custodial point, $y_1 = y_2$, the lightest singly charged and neutral particles must again combine to form an exactly degenerate $SU(2)$ multiplet. As  the mass of the doubly charged states stays constant, the only possibility is that the LNP is in a $3_M$, in agreement with what is observed Fig.\ref{fig:3M4D}. 
 Within the same approximations as above we get that, around  $y_1 = y_2$, the mass splitting between the charged component and the LNP is  again
\begin{equation}
\Delta m \approx {1\over 24} (y_1 - y_2)^2 {v^2\over m_M} > 0
\end{equation}
At the point $y_1 = y_2$ the doubly charged states belong to a $5_M$, but this multiplet does not contain the LNP.

\item  \underline{$m_D \gg m_M$} The main difference compared to $m_D \ll m_M$  is that the mass splittings are parametrically smaller. From inspection of the right panel of Fig.\ref{fig:3M4D}, we see that the LNP is part of $3_M$  for all the range of Yukawa couplings; this multiplet is essentially the original Majorana triplet.  Near $y_2 = \pm  y_1$, and for  $m_D \gg m_M \gg y_{1,2} v$, we get
\begin{equation}
\Delta m  \approx {1 \over 9} (y_1^2 - y_2^2)^2 {v^4 \over m_D^3} > 0
\end{equation}
 We see that the mass splitting is indeed parametrically smaller than in the case $m_D \ll m_M$ as it involves four powers of the Higgs vev, compare with Eq.(\ref{eq:split2}). The mass splittings away from the custodial points depend too on the hierarchy of Majorana and Dirac masses, a feature already observed in \cite{Tait:2016qbg}. This is illustrated diagrammatically in Fig.\ref{fig:massSplits} for $m_M \gg m_D$  (left panel) and $m_M \ll m_D$ (right panel). These Feynman graphs mean to illustrate the fact that mass splitting within custodial $SU(2)$ multiplets requires both  $y_1\neq y_2$  and a Majorana mass insertion. 
\end{enumerate}

To recap, in the $3_M 4_D$ system, at the custodial points, the pattern of multiplet is as in (\ref{eq:pattern}), with $3_M 4_W 4_W \rightarrow  3_M 3_M 5_M$. Whether the LNP is in a $5_M$ or a $3_M$ depends on the hierarchy between $m_M$ and $m_D$, as summarized in Table \ref{tab:charged}. 

%

%

\bigskip

\paragraph {The $3_M 2_D$ and $5_M 4_D$ cases} have common features. The spectra of the neutral states are analogous to those of the $1_M 2_D$ and $3_M 4_D$ systems. The main difference is that all states (neutral, charged and, if they exist, doubly charged) mix, see Figs.\ref{fig:3M2D} and \ref{fig:5M4D}.  Again, we distinguish $m_M > m_\ast \sim m_D$ and $m_M < m_\ast$.

\begin{enumerate}
\item\underline{$m_M \gg m_D$}  At the custodial point $y_2 = - y_1$, the LNP is the combination of Weyl states $\psi$ and $\tilde \psi$ that does not couple to the Higgs, and so has mass $m_D$. It is a $1_M$ in the $3_M 2_D$ case (Fig.\ref{fig:3M2D}), and is in a $3_M$ in the $5_M 4_D$ one (Fig.\ref{fig:5M4D}). 
Away from $y_1 = - y_2$, level repulsion decreases the mass of the LNP. Interestingly, because all the states are mixed, we see in the left panel of Fig.\ref{fig:3M2D} (Fig.\ref{fig:5M4D}) in the $3_M 2_D$ (resp. $5_M 4_D$) also the mass of 
the singly charged states $\chi_l^\pm$ (resp. doubly charged $\chi_l^{\pm\pm}$)  decrease, so that  at the other custodial point, $y_1 = y_2$, the LNP belongs to a $3_M$ (resp. a  $5_M$).
\item \underline{$m_M \ll m_D$} In this case, shown in the right panel of Fig.\ref{fig:3M2D} (Fig.\ref{fig:5M4D})  the LNP is always in a $3_M$ (resp. $5_M$) in the $3_M2_D$ (resp. $5_M 4_D$), as it is essentially the original Majorana $\chi_0$ with a small (in the limit $m_M \ll m_D$)  admixture of  $\psi_{1,2}^0$ states. 
\end{enumerate}

To recap, in the $3_M 2_D$ ($5_M 4_D$) system, and at the custodial points, the pattern of multiplet is as in (\ref{eq:pattern}), with $3_M 2_W 2_W \rightarrow  3_M 3_M 5_M$ (resp. $5_M 4_W 4_W \rightarrow  5_M 5_M 3_M$). Whether the LNP is in a $1_M$ or a $3_M$ (resp. a $3_M$ or a $5_M$) depends on the hierarchy between $m_M$ and $m_D$, see Table \ref{tab:charged}. 
  
\subsubsection{Comments on effects of loop corrections}
The conclusions of the previous section raises the question of the effects of radiative corrections. The custodial symmetry is broken at one-loop by electroweak corrections. For pure MDM, $\Delta m \propto \alpha_2 m_W \sin^2 \theta_W  = {\cal O}(100)$ MeV ~\cite{Cirelli:2005uq}. For mixed states, one expects that the situation is more complex. We have not studied the spectra at one-loop, so we will be sketchy, but we may refer to other works.  

A first naive conclusion would be that, at the custodial points, as
the LNP belongs to a multiplet of $SU(2)$, the situation must be the
same as for MDM. That this is not quite the case is illustrated in
Fig.~2 of ref.~\cite{Tait:2016qbg} for the $3_M 4_D$ case when
including NLO corrections. Beware that we used different conventions,
so their case $y = y_1 = y_2$ corresponds to our case $y = y_1 =
-y_2$. Regardless, their Fig.~2, illustrate the mass splittings
dependence in $y$, at one-loop, at one of the custodial points. The LNP
is noted $\chi_1^0 \equiv \chi_l^0$ and at tree level it is in a $5_M$
if $m_M \gg m_D$ and a $3_M$ if $m_M \lesssim m_D$ (see our
Fig.~\ref{fig:3M4D}).  One first sees in their Figure 2 that the mass splitting between the LNP and its singly charged partners
depends on $y$. This is manifest for $m_M \gtrsim m_D$ (left panel), in which
case the LNP has mass $m_D$ and is a Majorana built of the states
$\psi$ and $\tilde \psi$. As these states have opposite hypercharge,
their coupling to the neutral gauge bosons breaks the custodial
symmetry even if at the custodial point $y_1 = y_2$. For the case $m_M
\lesssim m_D$ (right panel), the DM is essentially the original Majorana multiplet, with
an admixture of Weyl states, so we expect this case to be closer to
MDM. The dependence on $y$ must be mild, consistent with the right
panel of Figure 2 of ref.~\cite{Tait:2016qbg}.

\begin{figure}[t]
  \begin{center}     
      \includegraphics[width=7.5cm]{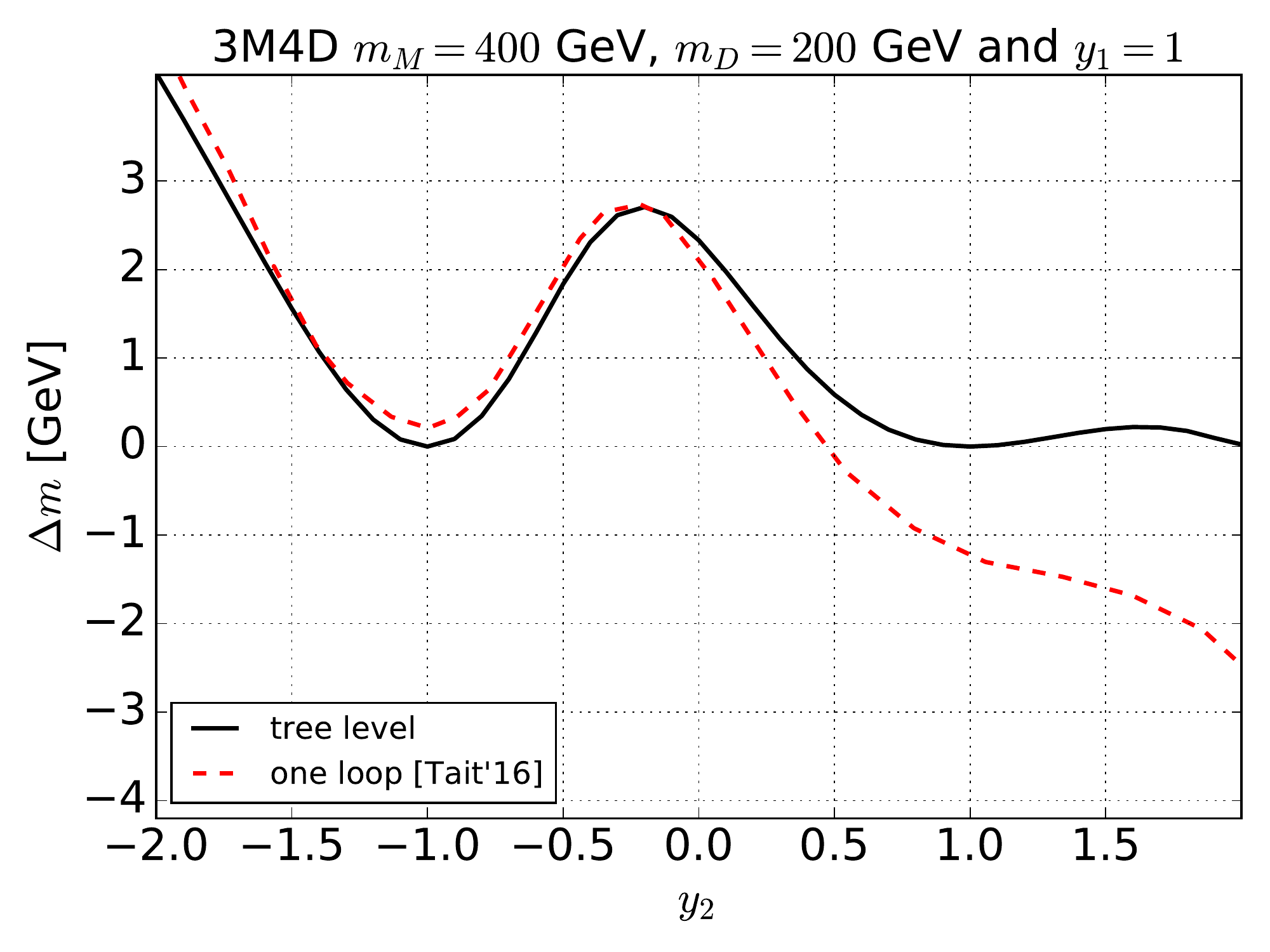}
  \end{center}
  \caption{Mass splitting $\Delta m = m_{\chi_l^\pm} - m_{\chi_l^0}$
    in the $3_M4_D$ system at tree level for $y_1=1$ as a function of
    $y_2$ for $m_M \gtrsim m_\ast$. The continuous black curve is the
    mass splitting at tree level. The red dashed curved is that at
    one-loop as obtained in ref.~\cite{Tait:2016qbg}. }
    \label{fig:split3M4D}
\end{figure}

Another naive conclusion would be that, away from the custodial
points, the LNP remains the lightest component of the multiplet even
at one-loop. After all, in the MDM, radiative corrections make the
charged partners heavier than the neutral one. However, it seems that
this is not the case either, see again ref.~\cite{Tait:2016qbg}. To be
precise, if we remain in a regime in which the Yukawa couplings are
not ``too large", one may expect that the dominant contributions to
mass splitting are either determined from $\vert y_1 \vert \neq \vert
y_2\vert$ at tree level or at one-loop through gauge corrections; in
both cases, the mass splittings are such that the LNP must be the
lightest stable particle and thus potentially a dark matter
candidate. If the Yukawa couplings get large however, this intuition
may become invalid. For instance, one may get into a regime in which
the mass of the LNP (and its charged partners) vanishes at tree
level. This is possible if $y_1$ and $y_2$ are large and have the same
sign (again, following our convention), see our
Fig.~\ref{fig:3M4D}. More precisely, one may check that this occurs if
the product $m_M m_D \approx \eta^2 y_1 y_2 v^2$, so that it may
happen only for bare Majorana and Dirac masses below the TeV range
provided $y_{1,2} \lesssim 4\pi$.  For the sake of comparison, we show
in Fig.~\ref{fig:split3M4D} both the mass splitting at tree level
derived here (black curve) and the result at one loop obtained in
ref.~\cite{Tait:2016qbg} (we report here with red dashed line the red
curve ref.~\cite{Tait:2016qbg} plotted the left panel of their
Fig.~4). There we see that $\Delta m$ at one loop (red dashed) becomes
negative when $y_1$ and $y_2$ are large and have the same sign,
corresponding to the range of parameters for which the mass of the
components of the lightest multiplet, and their mass splittings, are driven to zero at tree level
(black continuous). That one-loop corrections can jeopardize the mass
splitting in these conditions is thus perhaps not surprising. More
strange is the fact, stated in ref.~\cite{Tait:2016qbg}, that $\Delta
m$ becomes negative at one-loop even if the bare masses are large,
which we suppose corresponds to $m_M m_D \gg \eta^2 y_1 y_2
v^2$. Also, ref.~\cite{Tait:2016qbg} reports that this happens for
$m_M \gtrsim m_D$. It could be interesting to explore further this
feature.

\section{HMDM: cosmology and astrophysics}
\label{sec:dark-matt-phen}

The questions that we would like to address now is what is the mass
range for which our candidates can accommodate all the DM
(i.e. $\Omega_{DM}h^2=0.12$) and where, within this mass range, one
would expect to get observable signals from the dark matter? As
mentioned in the introduction, a complete treatment of these questions
would require to take into account Sommerfeld corrections and bound
state formation contribution to the annihilation cross-section for
arbitrary Majorana-Dirac mixing. This is a difficult problem, which
has only been tackled in details for specific SUSY-inspired scenarios,
see e.g.~\cite{Beneke:2016jpw,Beneke:2016ync,Beneke:2014gja}. It is
beyond the scope of this work to discuss these non-perturbative
corrections in the generic HMDM. In what follows, we first analyze the
viable parameter space in the perturbative limit. We then review how
non-perturbative corrections affect these predictions for the limiting
cases of pure MDM, and we provide an estimate of the Sommerfeld
corrections for the pure quadruplet scenario. The latter is the only
MDM case for which the Sommerfeld effect has not yet been explicitly
studied in the literature. We close the discussion on non perturbative
effects deriving the boundaries of the parameter space of the viable
HMDM under study in this paper making use of the $SU(2)_L$ symmetric
limit.  Finally, we briefly comment on the possible prospects for DM
direct and indirect searches. As we focus on candidates in the
multi-TeV range, collider searches are not relevant and are altogether
ignored in our discussions.\footnote{ See
e.g.~\cite{Ostdiek:2015aga,Garcia-Cely:2015quu,Arina:2016rbb,Cirelli:2014dsa,Xiang:2017yfs,Ismail:2016zby,DelNobile:2015bqo}
for recent MDM collider prospects related analysis.}

\subsection{HMDM  enlarging the MDM space: perturbative results }
\label{sec:DM-viab}

In this section we want to explore  to which extent the parameter
space of Minimal Dark Matter candidates is enlarged when different multiplets are coupled to the Higgs. This of course has been discussed case by case in many works, but as far as we know, no systematic comparison has yet  been provided in the literature.  
For a given system, say the $3_M 2_D$, the parameter space is {\em a
  priori} 4 dimensional, as we have two bare masses, $m_M$ and $m_D$
and two Yukawa couplings, $y_1$ and $y_2$. Fixing the relic abundance
reduces this to 3 independent parameters (the ``viable" DM
candidates). For pure MDM, and thus zero Yukawa couplings, the mass
of the viable DM candidate is fixed \cite{Cirelli:2005uq} and for
non-zero Yukawa couplings, the viable candidates should cover a domain
in the plane $m_M-m_D$.

To {\it estimate the boundary} of the HMDM
domains, we will make use of the electroweak symmetric limit. We will
do so first because this tremendously simplifies the discussion, as we
may neglect the mass splittings, mixing effects and annihilation
through Higgs mediated processes in determining the abundance. A
further motivation is that we may expect that the boundaries
correspond to candidates for which Yukawa couplings are small, and so
are close to the pure MDM cases. Last, the masses of MDM candidates are
typically in the multi-TeV range,  at least for MDM multiplet
larger than the doublet, so that freeze-out occurs close or above the
electroweak phase
transition~\cite{Cirelli:2005uq,Cirelli:2009uv}. Nevertheless, we
should keep in mind that the symmetric approximation is better for the
largest multiplets we consider.~\footnote{Concretely, the electroweak
  symmetric limit is expected to be most appropriate when DM
  interactions freeze-out at a temperature above the Electroweak Phase
  Transition (EWPT).  Assuming that the critical temperature at which
  $SU(2)_L$ gets restored is of $T_{cr}= 155$ GeV, the $SU(2)_L$
  symmetric limit would be expected to begin to be accurate for
  $m_{DM}\gtrsim x_f\times T_{cr} \sim 3$ TeV. Notice though that, in
  e.g. the case of the triplet DM with $m_{DM}=2.7$ TeV, the $SU(2)_L$
  symmetric limit Sommerfeld correction gives an estimate of the DM
  mass that is only $\sim 10\% $ larger than the one obtained in the broken limit,
  see~\cite{Mitridate:2017izz}. } We will comment further on the
validity of this approximation towards the end of this section.

\begin{figure}[h!]
  \begin{center}     
      \includegraphics[width=12cm]{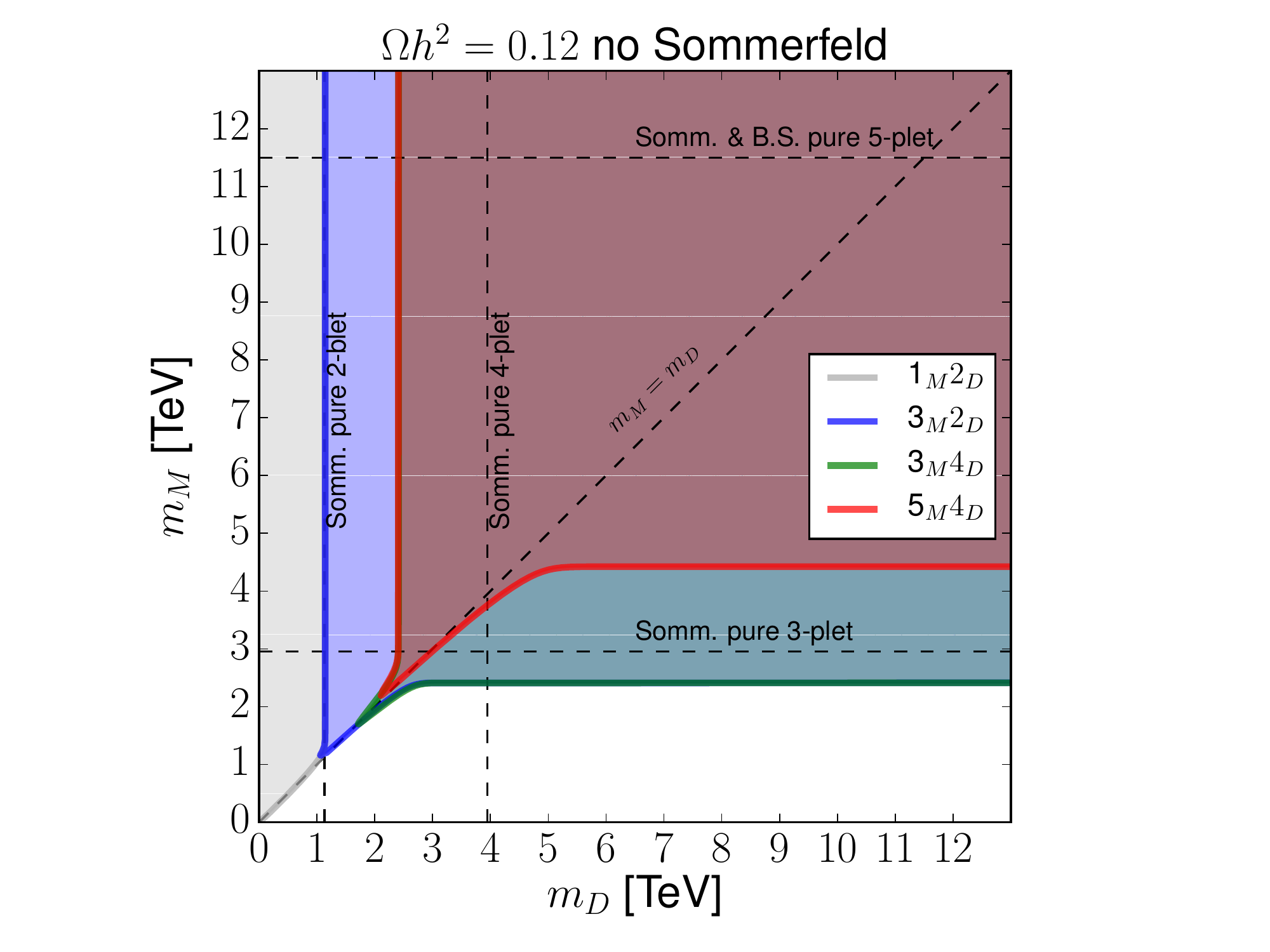}
  \end{center}
  \caption{ Viable parameter space in the perturbative $SU(2)_L$ symmetric
    approximation. Shaded regions enclose all models giving rise to
    $\Omega h^2=0.12$ for the 1M2D (in gray), for the 3M2D (in blue),
    for the 3M4D (in green) and 5M4D (in red). The limits of these
    contours, represented with thick continuous lines, have been
    obtained in the pure gauge limit, see text for details. The pure
    doublet, triplet, quadruplet and quintuplet limits including
    Sommerfeld corrections (and bounds state (B.S.) formation
    from~\cite{Mitridate:2017izz} in the 5-plet case) are indicated
    with dashed lines, see Sec.~\ref{sec:DM-phen-beyond}. }
\label{fig:pert}
\end{figure}

In the symmetric limit, we may neglect the mass splittings between the
multiplet components, so that our ingredients are a mixture of pure
Dirac and Majorana multiplets, which may co-annihilate with each
other if their masses are within $\sim 10\%$~\cite{Griest:1990kh}. On the other hand, in the presence of
Yukawa interactions between the Dirac and the Majorana multiplets one
expect that for $m_M\simeq m_D$ the coannihilation processes are quite
efficient. To determine the boundary of the HMDM
domains,  we assume that the Yukawa couplings are sufficiently large for
co-annihilations to be relevant,
but that they are small enough so that  the
DM $n$-plet annihilation cross-section  relevant for freeze-out is dominated by gauge interactions: 
\begin{equation}
\sigma v_{{\rm eff}, n}\simeq \frac{\zeta}{n^2}\frac{\alpha_2^2 {\cal C}_n}{m_{DM}^2}  
\label{eq:svsym}
\end{equation}
where $\zeta=1$ for the Majorana multiplet and 1/2 for the Dirac one,
and ${\cal C}_n$ is a dimensionless coefficient that mainly depends on
$n$ (see Sec.~\ref{sec:mixed-dm:-sommerfeld} below for more
details). Also, we have neglected the mass of the gauge
bosons. Following the treatment of~\cite{Griest:1990kh}, a proxy for
the total annihilation cross-section at freeze-out for a mixture of
Dirac and Majorana multiplets in interaction, would be:
\begin{eqnarray}
  \sigma v_{\rm eff}&\simeq&\frac{1}{g_{\rm eff}^2}\sum_{i=M,D} g_i^2\sigma v_{{\rm eff}, i} \quad {\rm and}\quad g_{\rm eff}=\sum_{i=M,D}g_i\cr
   g_i &= &n_i (1+\Delta_i)^{3/2} \exp(-\, x_f \Delta_i)
\label{eq:svefflim}
\end{eqnarray}
where the sum runs over the two multiplets and
$\Delta_i=(m_i-m_0)/m_0$ with $m_0=\min(m_M,m_D)$, $n_i$ denotes the
total number of degrees of freedom for the Majorana ($M$) or Dirac
multiplet ($D$) and $\sigma v_{{\rm eff}, i}$ corresponds to (\ref{eq:svsym}) for
$n=n_i$. For concreteness, we will take $x_f= m_0/T_f= 30$ when computing the cross-sections
in the $SU(2)_L$ symmetric limit.  We  also use the standard approximate expression for the relic abundance
\begin{eqnarray}
\label{eq:relic}
  \Omega_{\rm DM}h^2&\simeq&\frac{1.07\, 10^9 \, x_f}{M_{pl}/{\rm GeV}\sqrt{g_*}\, \eta\,  \sigma v_{eff}}\,,\label{eq:om}
\end{eqnarray}
valid for annihilation into an s-wave, with $M_{pl}= 1.22 \, 10^19 $ GeV is the Planck mass and $g_*$ is the number of relativistic degrees of freedom at the time of freeze-out. Imposing $\Omega h^2=0.12$, we obtain 
the contours shown in Fig~\ref{fig:pert} with continuous colored
lines. Notice that the material necessary to work out the expression
of the relevant annihilation cross-sections is discussed in more
detail in Sec.~\ref{sec:DM-phen-beyond}.

For each pair of Dirac and Majorana multiplets, the contours have asymptotic solutions
corresponding to the pure (Majorana or Dirac) MDM candidates, linking each
others approximatively along the diagonal $m_M= m_D$. Along this diagonal, 
the effective number of
degrees of freedom is larger than for the pure cases, an effect which must be compensated by larger annihilation cross sections and thus   smaller DM masses, compared to the pure cases. To put it simply, the situation is like having together two DM particles, with a similar mass, and so a larger abundance for fixed annihilation cross sections. 
This is the origin of the bottom-left pointing nose-shaped features
observed in the contours along the $m_M \sim m_D$ direction. For
larger Yukawa couplings DM depletion is more efficient due to the
opening of more annihilation channels and more efficient
co-annihilation channels, and so with extra terms contributing to
eq.~\ref{eq:svefflim}, see \cite{Griest:1990kh}. Thus the contours
feature a top-right pointing ``nose'' instead, {\em i.e.} the observed
relic abundance would be obtained for a value $m_M=m_D$ larger than
for the pure cases. Such features are observed in the plots of
Ref.~\cite{Tait:2016qbg} for the case $3_M4_D$.  Thus we infer that
the shaded regions delimited by the contours (gray for 1M2D, blue for
3M2D, green for 3M4D and red for 5M4D) enclose all the candidates that
would give rise to $\Omega h^2=0.12$ for a proper choice of the
Yukawas $y_1,y_2$.  For a given model, larger couplings are required
in the innermost regions when larger $(m_D,m_M)$ masses are
considered. Outside the shaded regions, the DM candidates have an
abundance below $\Omega h^2=0.12$.

\begin{figure}[t!]
  \begin{center}     
      \includegraphics[width=12cm]{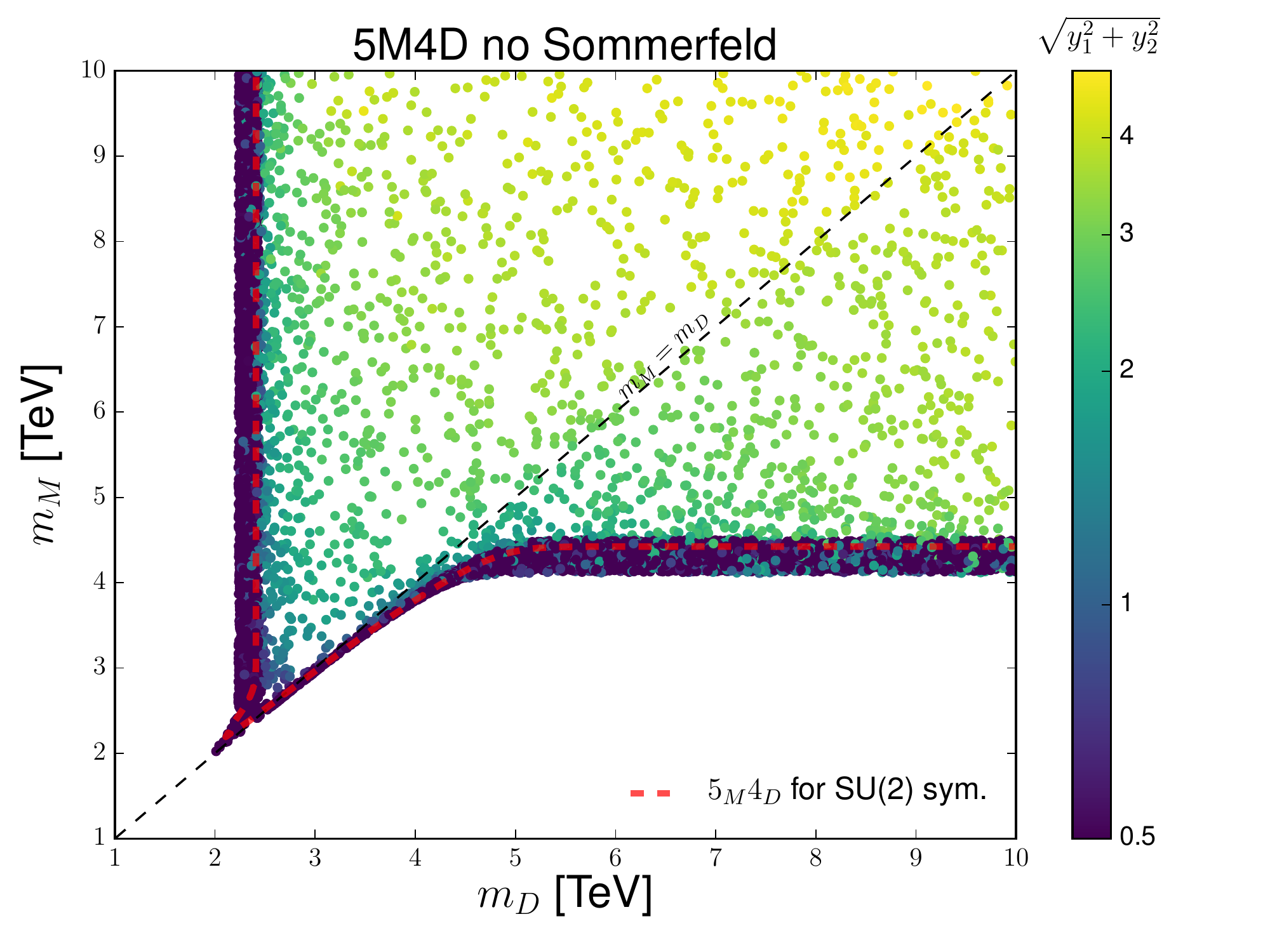}
  \end{center}
  \caption{ Viable parameter space in the perturbative 5M4D case for
    an explicit integration of the dark matter abundance with {\tt
      micrOMEGAs} in the $SU(2)_L$-broken limit. All points give rise to
    $\Omega h^2\simeq 0.12$ for a value of the yukawa combination $
    \sqrt{y_1^2+y_2^2}$ indicated with the color code. Notice that we
    have considered $|y_1|,|y_2|$ as small as $10^{-4}$ but all points
    with $\sqrt{y_1^2+y_2^2}< 0.5$ are shown in blue as they all end
    up in the contours of the 5M4D parameter space.  With red dashed
    line, we show the (red) contour obtained in the $SU(2)_L$ symmetric
    limit for the 5M4D case in Fig.~\ref{fig:pert}.}
\label{fig:pert5M4D-yeff}
\end{figure}

To corroborate this simple, yet qualitative picture we have checked
that the contour, obtained here in the electroweak symmetric limit, is
in a good agreement with the numerical results for the dark matter
abundance computed with {\tt micrOMEGAs}, i.e. working in the
$SU(2)_L$ broken limit, including mass splittings. For illustrative
purposes, we show in Fig.~\ref{fig:pert5M4D-yeff} the results from a
random scan over the parameter space of the $5_M 4_D$ system, imposing
$0.11<\Omega h^2<0.13$, $ 10^{-4} <|y_1|,|y_2|< 4\pi$ and
$1.5<m_\chi,m_\psi< 10 \mbox{TeV}$. Let us emphasize that we do not
incorporate the possible non perturbative effects in
Fig.~\ref{fig:pert5M4D-yeff}. The latter effects are discussed in the
next section.  Yet, we see that the viable parameter space of
candidates obtained with {\tt micrOMEGAs} (colored points) fit very well
within the boundaries obtained in the $SU(2)_L$ symmetric limit, shown
with dashed red contour (corresponding to the continuous red colored
line in Fig.~\ref{fig:pert}). The latter was obtained using the simple
equations ~(\ref{eq:svefflim}) and (\ref{eq:relic}).

\subsection{Dark matter abundance and Sommerfeld corrections}
\label{sec:DM-phen-beyond}

As mentioned above, computing Sommerfeld corrections in each HMDM case
in general is a very involved calculation.  In the $SU(2)_L$ symmetric
limit, important simplifications of the Sommerfeld computation come
from the fact that isospin is conserved in the annihilation and
scattering processes.  This allows to solve Schrodinger equations of
2-particle wavefunctions $\Psi_{I}$ of definite total isospin $I$,
without mixing among them. As a consequence the Sommerfeld correction
compution of a system of a large number $N$ of coupled differential
equation is reduced to the resolution of $N'<N$ {\it uncoupled}
differential equations, which strongly simplifies the
problem~\cite{Strumia:2008cf,
  deSimone:2014pda,Garcia-Cely:2015khw,Garcia-Cely:2015dda,Garcia-Cely:2015quu,Asadi:2016ybp,Mitridate:2017izz}.
We will work in this framework in what follows.

\subsubsection{Sommerfeld corrections in the $SU(2)_L$ symmetric limit}
\label{sec:somm-corr-unbr}

The $N'$ above is associated to the number of possible irreducible
representations ${\cal R}_a$ resulting from the direct product:
\begin{equation}
R_i\otimes R_j=  \sum_{k=1}^{N'}{\cal R}_a  
\end{equation}
where $R_i$ and $R_j$ denote the representation under $SU(2)_L$ of the
two annihilating particles $i$ and $j$. Assuming zero mass gauge
bosons in the unbroken $SU(2)_L$ limit, the potentials driving the
$SU(2)_L$ long range interactions, take the
form~\cite{deSimone:2014pda}:
\begin{equation}
  V_{I_a}^{SU(2)}(r) =\frac{\alpha_{I_a}}{r}=\frac{\alpha_2}{r}\frac12 \left(C_a-C_i-C_j\right) \,,
\label{eq:VSU2}
\end{equation}
where $\alpha_2=g/4\pi$, with the $SU(2)_L$ gauge coupling $g$, and
the $C_l$ with $l= i, j$ and $a$ are the quadratic Casimir operators
associated to the representation $R_i$,$R_j$ and ${\cal R}_a$. In the
case of $SU(2)_L$, $C_l=I_l(I_l+1)$ where $I_l$ is the isospin
corresponding to the representation $R_l$. Also, for annihilating
particles with non zero hypercharge, we get a U$(1)_Y$ contribution to
the potential that reads:
\begin{equation}
  V^{U(1)}=\frac{\alpha'}{r} = \frac{-\alpha_2t_w^2Y^2}{r}
\label{eq:U1}
\end{equation}
where $\alpha'= g'/(4 \pi)$, $g'$ is the U(1)$_Y$ gauge coupling
related to $g$ by the tangent of the Weinberg angle $t_w$ and $Y= |Y_i|=
|Y_j|$ is the absolute value the hypercharge of the particles $i$ and
$j$.

In this way the total potential associated to a pair of particles
annihilating in the total isospin state $I=I_a$ becomes
\begin{equation}
V_I = V_I^{SU(2)} + V^{U(1)} = \frac{\alpha_I + \alpha'}{r}~.
\label{pottotal}
\end{equation}
In the zero mass approximation for the gauge bosons, each of the $N'$
Shr\"odinger equations can be solved analytically. As a result, in the
s-wave limit, the annihilation cross section $\sigma v_I$ of a given
total isospin $I$ 2-particles state  is given by:
\begin{eqnarray}
\sigma v_I=  S_I\, \sigma v_I^{\rm pert} \quad {\rm with}\quad S_I= \df{-\pi a_I}{1-\exp(\pi/a_I)} 
\label{eq:SI}
\end{eqnarray}
 where $S_I$ is the Sommerfeld factor that multiplies the perturbative
 annihilation cross section $\sigma v_I^{\rm pert}$ and $a_I
 =v/[2(\alpha_{I}+\alpha')]$ where $v$ denote the relative velocity of
 the initial state particles. A priori, one should be concerned with
 the fact that at finite temperature, the gauge boson masses are non
 zero. The Higgs vev is temperature dependent and, in addition, the
 squared masses of the gauge bosons get an extra thermal mass
 contribution, see e.g.~\cite{Cirelli:2007xd}. We have however checked
 that due to these effects, for large representations, the Sommerfeld correction factors obtained
 resolving the Shr\"odinger equations including the thermal mass
 corrections agree with the Coulomb approximation of eq.~(\ref{eq:SI})
 with an error $<1\%$ for $I\leq 2$ that is the maximum total isospin of a
 pair of standard model particles $XX'$ into which $ij$ is annihilating
 into. See also \cite{Mitridate:2017izz} for a careful treatment.

For computing the relic abundance in a pure case, we use eq.~(\ref{eq:om}) with
\begin{equation}
  \sigma v_{eff}= \zeta\sum_{ij} \df{g_ig_j}{g_{eff}}\sigma v_{ij} \label{eq:sveffij}
\end{equation}
with $\zeta=1$ for self-conjugate particles and $1/2$ otherwise and
$g_{eff}=\sum g_i$ with $g_i$ the number of degrees of freedom
associated to the species $i$. Notice that the eq.~(\ref{eq:sveffij})
is only valid in the limit of negligible mass splittings between the
(co-)annihilating particles that is relevant in the $SU(2)_L$ unbroken
limit.
The (co-)annihilation cross-sections of initial state particles
$ij$ to any 2-body SM final state, $\sigma v_{ij}$, can easily be
obtained from Feynmman rules. Making use of Clebsch-Gordan decomposition
one can recast the $|ij\rangle$ contributions in terms of the isospin
of 2 particle states $|I_a \rangle$, see appendix~\ref{sec:svij2svI}
for one example in the quadruplet case that is addressed in more detail
below. As a result, for a dark matter candidate in a
representation $R_X$ of $SU(2)_L$ with an isospin $I_X$ , in the
simple case of $Y=0$, the effective cross section of eq.~(\ref{eq:sveffij}) reduces to:
\begin{equation}
  \sigma v_{eff}= \frac{\zeta}{(2 I_X+1)^2}\sum_{I} (2 I+1)\, \sigma v_{I}\qquad [{\rm case} \, Y=0]\,,
\label{eq:sveffI}
\end{equation}
where $I$ runs over the $I_a$ values with $a=1,..,N'$. The
cross-sections $\sigma v_{I}$ should be taken as in
eq.~(\ref{eq:SI}).  For $Y\neq 0$, extra contributions to $\sigma
v_{eff}$ are expected from $U(1)_Y$ gauge bosons ($B_\mu$) insertions
giving rise to annihilation cross sections proportional to
${\alpha'}^2$, denoted by $\sigma v_{g'}$, and cross sections
proportional to $ \alpha'\alpha$, denoted by $\sigma v_{g'g}$. The
former results from $B_\mu$ mediated annihilations into two fermions
or two Higgs, corresponding to $SU(2)_L$ singlet state, while the
latter results from annihilations into both $B_\mu$ and an $SU(2)_L$
gauge boson, corresponding to $SU(2)_L$ triplet state. The overall
Sommerfeld-corrected effective cross section relevant for the relic abundance
computation  thus reads:
\begin{equation}
  \sigma v_{eff}=\frac{\zeta}{(2 I_X+1)^2}\left(\sum_{I} (2 I+1)S_{I}\sigma v_I^{\rm pert}+S_{I=1}\, \sigma v_{gg'}^{\rm pert}+S_{I=0}\,\sigma v_{g'}^{\rm pert}\right)\, \qquad [{\rm case} \, Y\neq0]\,,
\label{eq:sveffSIY}  
\end{equation}
where, in the sum, $I$ runs over the $I_a$ values with
$a=1,..,N'$. Let us emphasize that the perturbative results, used for
the plot in Fig.~\ref{fig:pert}, can simply be obtained setting  the
Sommerfeld  factors $S_I$ to 1. 

\subsubsection{One example: the pure quadruplet}
\label{sec:one-example:pure4}

We now illustrate in more detail how the method above can be applied
to the {\it pure} 4-plet dark matter case. To our knowledge, this is the
only pure case in which Sommerfeld corrections have not been
previously computed explicitly. The 4-plet appears in a study of
ref.~\cite{Tait:2016qbg}, a treatment at perturbative level only,
while the treatment of the doublet, the triplet, the quintuplet and
the 7-plet at non-perturbative level can readily be found in
refs.~\cite{Garcia-Cely:2015khw, Garcia-Cely:2015quu,
  Mitridate:2017izz,Asadi:2016ybp,Cirelli:2007xd,Cirelli:2015bda,Garcia-Cely:2015dda}. Our
results agree with the most recent updates, see
Sec.~\ref{sec:mixed-dm:-sommerfeld} for more details.

We thus provide here a detailed computation of the Sommerfeld
correction in the $SU(2)_L$ symmetric limit for the 4-plet. The
Weyl multiplets that we are dealing with are:
\begin{equation}\psi = \left(
\begin{array}{c}
\psi^{++} \\ \psi^{+} \\ \psi^0 \\ \psi^{-}
\end{array}
\right)\qquad {\rm and }\qquad
\tilde\psi = \left( 
\begin{array}{c}
\tilde\psi^{+} \\ \psi^{0} \\ \psi^- \\ \psi^{--}
\end{array}
\right),
\label{4plet}
\end{equation}
with opposite hypercharges equals to 1/2 and -1/2.  In the scattering
of 4 and $\bar 4$, we know that $4\otimes\bar 4 = \sum_{a=1}^{N'}
{\cal R}_a = 1\oplus 3 \oplus 5\oplus 7$, where ${\cal R}_a$ are the
$SU(2)_L$ representations of the 2-particle states with $a=1,..,4$ and isospins $I= \{0,1,2,3 \}$. In the Coulomb limit, the associated
$SU(2)_L$ potentials from eq.~(\ref{eq:VSU2}) take the values
\begin{equation} 
V^{SU(2)}_I = 
\df{-\alpha_2}{r}\left\{\frac{15}{4},\frac{11}{4},\frac{3}{4},\frac{-9}{4}\right\}\, , \qquad \mbox{ [4-plet]}
\label{potential}
\end{equation} 
where we have used that the 4-plet has isospin $I_{4}=3/2$. In
addition, the $U(1)_Y$ contribution reads
\begin{equation}
V^{U(1)}= -\alpha_2t_w^2Y^2_4/r \quad {\rm with} \quad Y_4=1/2 \,.  \qquad \mbox{ [4-plet]}
\end{equation}
The overall potentials for $I=
\{0,1,2,3 \}$ involved in the long range physics computation
associated to the annihilation of the 4 and $\bar 4$ is thus a sum of
$SU(2)_L$ potentials from eq.~(\ref{potential}) and $V^{U(1)}$ as in
eq.~(\ref{pottotal}). Using~(\ref{eq:SI}) with $v\simeq 0.2$,\footnote{We use
  $v=0.23$ for the computation of $S_{I_ a}$ so as to match the results
  of~\cite{Garcia-Cely:2015dda,Cirelli:2015bda} in the 5-plet case for
  which the Sommerfeld correction in the $SU(2)_L$ symmetric limit have
  been shown to provide an accurate approximation to the full
  computation~\cite{Mitridate:2017izz}.} we obtain the following Sommerfeld correction factors:
\begin{eqnarray}
S_{I} = \{3.9,3.0,1.5,0.3\}\,.  \qquad \mbox{ [4-plet]}
\end{eqnarray}

\begin{figure}[t]
  \begin{center}     
        \includegraphics[width=10cm]{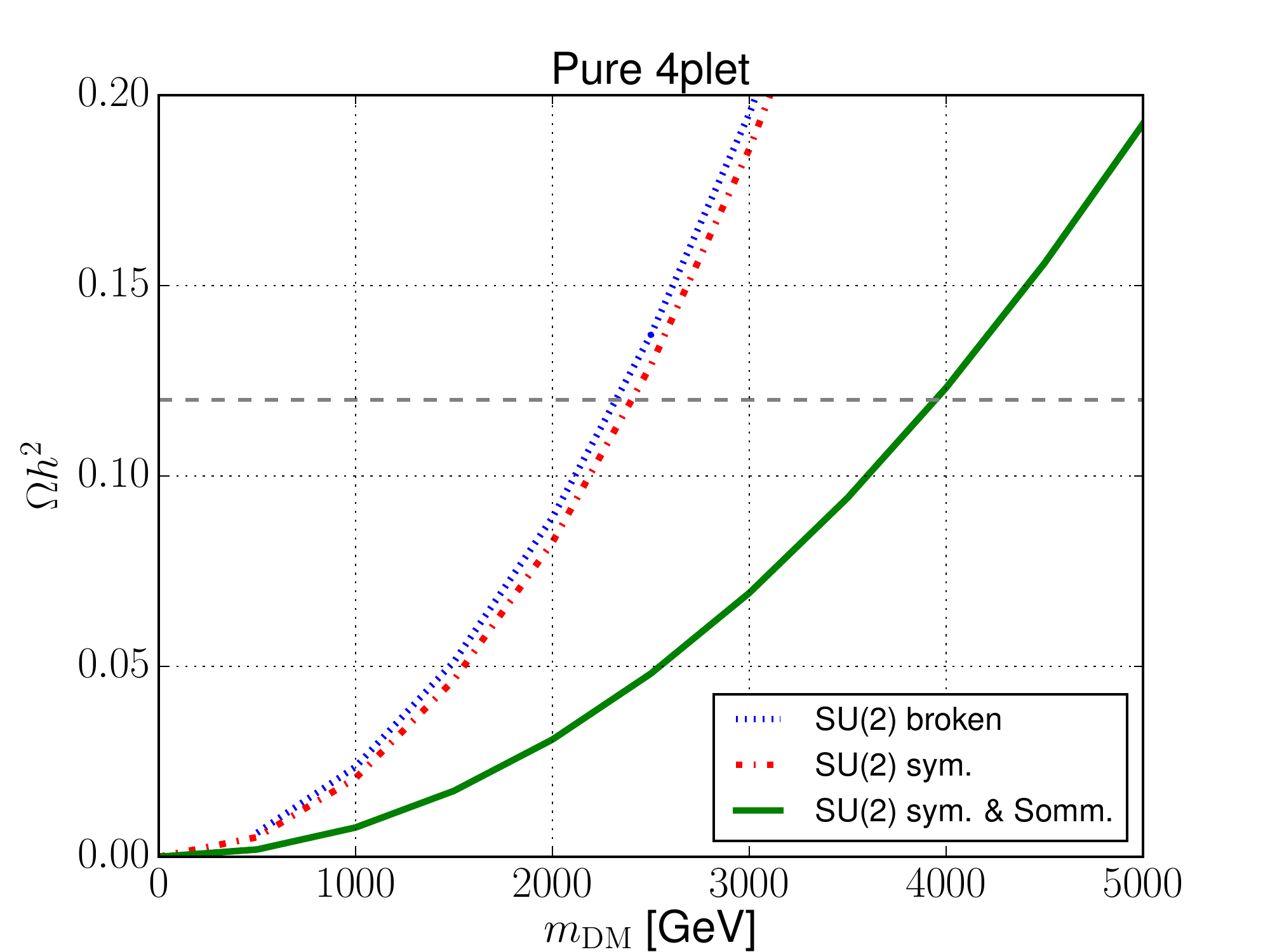}
          \end{center}
  \caption{Pure 4-plet relic abundance $\Omega h^2$  including the
    Sommerfeld corrections (continuous brown) or not (red dotted) in
    the s-wave $SU(2)_L$ symmetric limit for the annihilation of
    quadruplet dark matter. With the blue dashed line we also show the
    results obtained with {\tt micrOMEGAs} neglecting the Sommerfeld
    corrections in the $SU(2)_L$ broken case.  }
\label{fig:Om}
\end{figure}

After extracting the $\sigma v_{I,g,g'}$ following the method above,
see appendix~\ref{sec:svij2svI} for more details, the results for the
relic abundances in the s-wave $SU(2)_L$ symmetric limit are summarized in
Fig.~\ref{fig:Om} using:
\begin{eqnarray}
&&\sigma v_{I=0}^{\rm pert}=\frac{75}{4}\frac{\alpha_2^2\pi}{M_{DM}^2}, \quad
\sigma v_{I=1}^{\rm pert}=\frac{125}{8}\frac{\alpha_2^2\pi}{M_{DM}^2},\quad
\sigma v_{I=2}^{\rm pert}=6\frac{\alpha_2^2\pi}{M_{DM}^2} \\
&& \sigma v_{g'g}^{\rm pert}=\frac{15}{2}\, t_w^2\,\frac{\alpha_2^2\pi}{M_{DM}^2}, \quad \sigma v_{g'}^{\rm pert}=\frac{43}{8}\, t_w^4\,\frac{\alpha_2^2\pi}{M_{DM}^2} \quad\mbox{ [4-plet]}
\label{eq:sv4-plet}
\end{eqnarray}
From Fig.~\ref{fig:Om}, in order to account for $\Omega_{DM}h^2=0.12$,
one would thus get $M_{DM}= 2.4$ TeV in the perturbative limit, while
taking into account the Sommerfeld corrections one gets $M_{DM}= 3.9$
TeV. Also notice that, working in the $SU(2)_L$ broken limit using
{\tt micrOMEGAs} to compute the relic abundance, one obtains $M_{DM}=
2.3$ TeV in the perturbative limit to account for
$\Omega_{DM}h^2=0.12$ (see the blue dashed line in
Fig.~\ref{fig:Om}). This agrees with results of~\cite{Tait:2016qbg} in
the $3_M4_D$ case in the limit of high mass triplet (i.e DM almost
pure quadruplet). We are thus making a $\sim 4\%$ error working in the
$SU(2)_L$ symmetric case in order to determine the relevant dark
matter mass in the perturbative limit.

\vspace{.3cm}
It has recently been pointed out that bound state formation (BSF) can
provide an extra enhancement of the annihilation cross-section of
minimal dark matter~\cite{Asadi:2016ybp,Mitridate:2017izz}. In
particular~\cite{Asadi:2016ybp} first showed that the rate of BSF in the triplet case is suppressed compared to direct
annihilation. In~\cite{Mitridate:2017izz}, it was shown that BSF raises the mass of the 5-plet to 11.5 TeV, i.e. a
$\sim$ 20\% ($\sim$ 40\%) correction to the mass (annihilation cross
section) obtained with Sommerfeld corrections only while essentially
no corrections appear in the 3-plet case. 

It is beyond the scope of this paper to compute in detail the impact
of BSF on freeze-out calculations. Here we just want to argue that
the correction from BSF corresponding to the 4-plet case is expected to
be smaller than for the 5-plet case. As noted by
~\cite{Mitridate:2017izz}, bound states can efficiently form even at
temperatures $T\sim m_{\rm DM}/x_f$ larger than the corresponding bound state
binding energies, because the dissociation rate can be suppressed with
respect to naive expectations. Nonetheless, the intuition that smaller
$E_B/T_f$ ratios (i.e. binding energy to freeze-out temperature) lead
to smaller corrections from BSF remains valid, as shown in
\cite{Mitridate:2017izz} for the 3-plet case compared to the 5-plet
case. Indeed for the former, $E_B\lesssim 0.05$ GeV at $T_f\sim 100$
GeV leads to a correction to the DM relic density at the \% level,
whereas for the latter, $E_B\lesssim 60$ GeV at $T_f\sim 460$ GeV
leads to a 40\% correction. In the case of the 4-plet, the most
attractive potential (corresponding to the singlet two-particle state) has a strength of $15\alpha_2/4$, which corresponds to an
$n=0$ bound state with binding energy $E_B\sim 4.2$ GeV at $T_f\sim
160$ GeV, following the method of estimation of
~\cite{Mitridate:2017izz}. As can be noted, $E_B/T_f$ is a factor
$\sim 5$ smaller for the 4-plet than for the 5plet, thus the BSF
correction to the relic abundance in the case of the 4-plet should be
much less important.

\subsubsection{HMDM: Sommerfeld correction of the viable parameter space}
\label{sec:mixed-dm:-sommerfeld}

\begin{table}[t!]
  \begin{tabular}{|c|c|c|c|c|c|c|}
    \hline
    $n$ & $I_a$ & $\lambda_a$ & $S_{I_a}$ & $\sigma v_{I_a}^{\rm pert}$& $m_{DM}^{\rm pert}$ [TeV]&$m_{DM}^{\rm Som}$ [TeV]\\
    \hline
    \hline
2    & $0$ & $\frac{3}{4}+\frac{1}{4}t_w^2 $ &1.5 & $\frac{3 \pi  \alpha_2^2}{8 m_{DM}^2}$ 
&1.1&1.1\\[.2cm]
& $1$ & $-\frac{1}{4}+\frac{1}{4}t_w^2 $ &0.9 & $\frac{25 \pi  \alpha_2^2}{16 m_{DM}^2}$ 
& & \\[.2cm]
\hline \hline
3& $0$ & $2 $ &2.3 & $\frac{4 \pi  \alpha_2^2}{m_{DM}^2}$ 
&2.4&3.\\[.2cm]
& $1$ & $1 $ &1.6 & $\frac{25 \pi  \alpha_2^2}{4 m_{DM}^2}$ 
& & \\[.2cm]
& $2$ & $-1 $ &0.6 & $\frac{\pi  \alpha_2^2}{m_{DM}^2}$ 
& & \\[.2cm]
\hline \hline
4& $0$ & $\frac{15}{4}+\frac{1}{4}t_w^2 $ &3.9 & $\frac{75 \pi  \alpha_2^2}{4 m_{DM}^2}$ 
&2.4&3.9\\[.2cm]
& $1$ & $\frac{11}{4}+\frac{1}{4}t_w^2 $ &3. & $\frac{125 \pi  \alpha_2^2}{8 m_{DM}^2}$ 
& & \\[.2cm]
& $2$ & $\frac{3}{4}+\frac{1}{4}t_w^2 $ &1.5 & $\frac{6 \pi  \alpha_2^2}{m_{DM}^2}$ 
& & \\[.2cm]
& $3$ & $-\frac{9}{4}+\frac{1}{4}t_w^2 $ &0.3 & $-$
& & \\[.2cm]
\hline \hline
5& $0$ & $6 $ &5.9 & $\frac{60 \pi  \alpha_2^2}{m_{DM}^2}$ 
&4.4&9.3\\[.2cm]
& $1$ & $5 $ &5. & $\frac{125 \pi  \alpha_2^2}{4 m_{DM}^2}$ 
& & \\[.2cm]
& $2$ & $3 $ &3.1 & $\frac{21 \pi  \alpha_2^2}{m_{DM}^2}$ 
& & \\[.2cm]
& $3$ & $0 $ &1. & $-$ 
& & \\[.2cm]
\hline
\end{tabular}
  \caption{For the pure multiplet of dimension $n$, the Isospins of
    the relevant 2-particle states are given by $I_a$, the potentials
    are driven by the $\lambda_a=-(\alpha_{I_a}+\alpha')/\alpha_2$
    couplings and, using $\sigma v_{I_a}^{\rm pert}$ together with the
    appropriate $ \sigma v_{g, \,gg'}^{\rm pert}$ in the 2-blet,
    4-plet cases, one obtains $m_{DM}$ TeV for the dark matter mass
    including Sommerfeld corrections {\it only} in the $SU(2)_L$ symmetric
    limit ($m_{DM}^{\rm pert}$ is obtained without Sommerfeld
    corrections).  }
  \label{tab:svI}
\end{table}

%
\begin{figure}[t!]
  \begin{center}     
        \includegraphics[width=12cm]{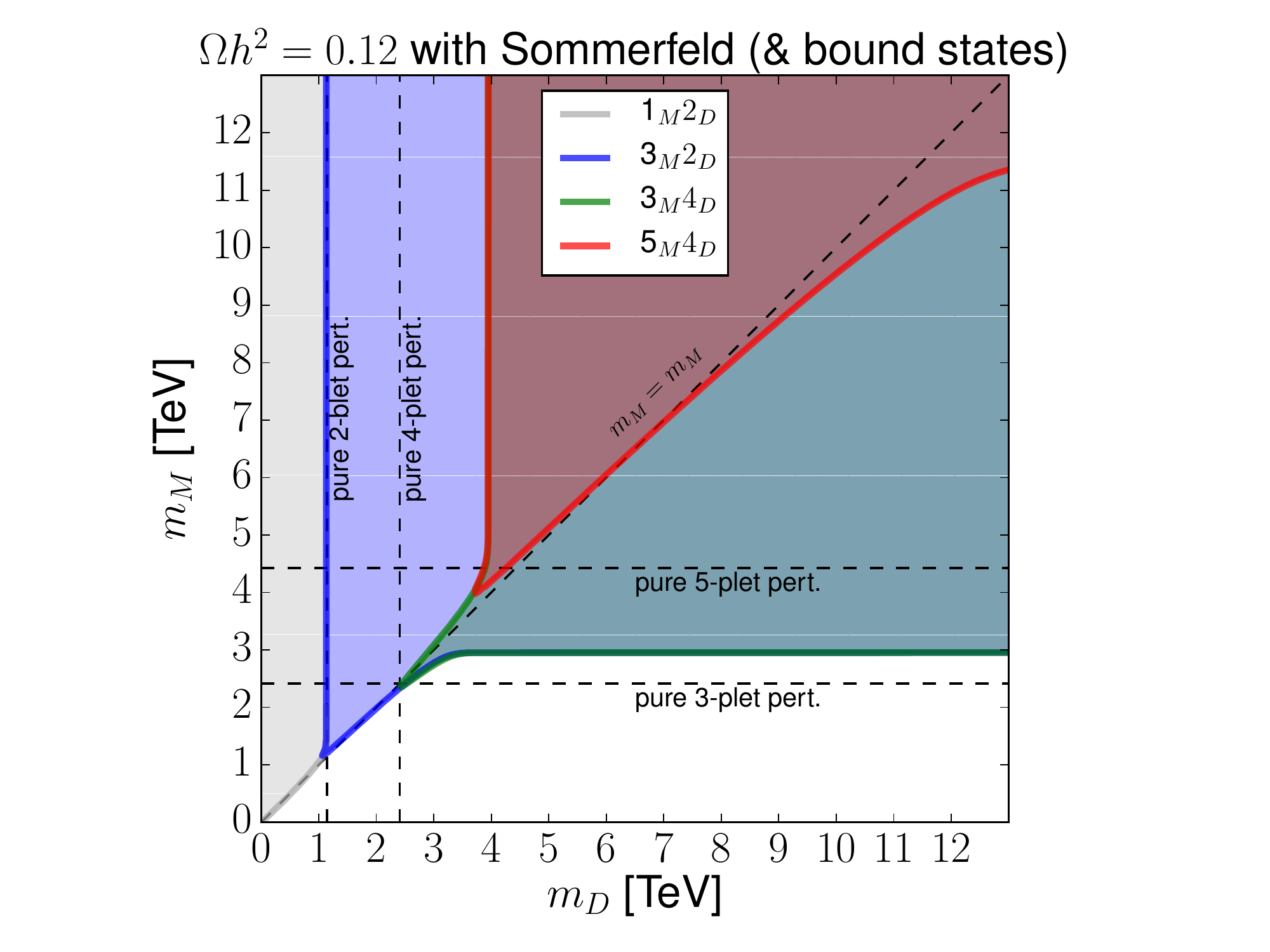}
    \end{center}
  \caption{Boundaries of the viable parameter space of HMDM models in
    the $m_M-m_D$ plane represented with continuous lines when
    computing the relic abundance in the $SU(2)_L$ symmetric limit
    with Sommerfeld corrections for the 2-blet, 3-plet, 4-plet
    limiting cases. For the 5-plet, BSF corrections
    of~\cite{Mitridate:2017izz} are taken into account. As a guide for
    the eye, the perturbative results for the pure MDM cases are indicated
    with dashed lines. }
\label{fig:som}
\end{figure}

The impact of Sommerfeld corrections on the viable space for dark
matter is illustrated in Fig.~\ref{fig:som}. In order to derive the
Sommerfeld enhanced pure n-plet limits we have followed the same
recipe as in the case of the 4-plet above. For all the pure cases,
corresponding to the limits $m_M\gtrsim (\lesssim) m_D$ of the models
considered here, we summarize our findings in
Tab.~\ref{tab:svI}. These results were obtained considering an average
velocity of $v\simeq 0.2$ in the computation of $S_{I_a}$ and the dark
matter masses for the candidate giving rise to all the DM assuming
$x_f=30$.  For the doublet, as in the case of 4-plet (see
eq.~(\ref{eq:sv4-plet}), one has to take into account $\sigma v_{g'}$
and $\sigma v_{gg'}$ (the $U(1)_Y$ and mixed $U(1)_Y$ \& $SU(2)_L$
contribution as in eq.~(\ref{eq:sveffSIY})).  In the s-wave limit, for
the doublet, we have found:
\begin{eqnarray}
&& \sigma v_{g'g}^{\rm
  pert}=\frac{3}{4}\, t_w^2\,\frac{\alpha_2^2\pi}{m_{DM}^2}\qquad
\sigma v_{g'}^{\rm pert}=\frac{43}{16}\,
t_w^4\,\frac{\alpha_2^2\pi}{m_{DM}^2} \quad \mbox{ [doublet]}  
\end{eqnarray}
 Also notice that in Tab.~\ref{tab:svI}, we only provide $\sigma
 v_{I_a}$ for $I_a<3$ as we focus on 2 body final states
 only which total isospin is always smaller than 3 in the SM. Our results are in
 agreement with the cases already available in the literature~\cite{Garcia-Cely:2015quu,Mitridate:2017izz}.


The dark matter mass obtained to match $\Omega h^2=0.12$ when
considering Sommerfeld corrections in the $SU(2)_L$ symmetric limit
are provided in the last column of Tab.~\ref{tab:svI} and can be
compared to the latest derived value present in the
literature. Considering Sommerfeld corrections only, one can get
from~\cite{Beneke:2014hja} $m_{DM}\simeq 1.2$ TeV in the doublet
case,\footnote{We extract the doublet case from
  ref.~\cite{Beneke:2014hja} in their Fig. 11 and table. 1 in the
  decoupling limit: $M_2> \mu$.} while in the 3-plet and in the 5-plet
case ref.~\cite{Mitridate:2017izz} reports $m_{DM}\simeq$ 2.7 TeV and
$m_{DM}\simeq$ 9.3 TeV respectively.  We see that the $SU(2)_L$
symmetric limit provides a very good way to {\it estimate} Sommerfeld
corrections at freeze-out. In the 5-plet case however, bound state
formation changes the dark matter annihilation cross-section and
eventually gives rise to the right abundance for $m_{DM}\simeq$ 11.5
TeV~\cite{Mitridate:2017izz}. We have not tried to re-evaluate this
effect here but we account for it in our summary plot of
Fig.~\ref{fig:som}.  In the latter plot, we make use of our results
from Tab.~\ref{tab:svI} except in the case of the 5-plet where we use
the BSF result from~\cite{Mitridate:2017izz}. The interpolating regions
between the pure cases have been obtained with the same method as in
the perturbative case, see Sec.~\ref{sec:DM-viab}, Eq.\ref{eq:svefflim}.

\subsection{Dark matter detection prospects}
\label{sec:dark-matt-detect}

As regards prospects for DM detection, we hereby discuss the main features and effects that can be expected when moving from the pure MDM scenarios to the HMDM ones, without providing a full-fledged analysis that would also require computing the conditions for the right relic abundances of the various HMDM scenarios.

\subsubsection{Direct detection}
\label{sec:direct-detect-search}

HMDM has spin-dependent and spin-independent interactions at tree
level with quarks. As mentioned in the introduction, we have checked
numerically that spin-dependent cross-section (computed at tree-level)
always appear to be way beyond the reach of current experiments, we
will thus focus here on spin independent (SI) scattering. For
the latter, the relevant processes for HMDM are scatterings with quarks via Higgs exchange  at
tree level and, at loop
level, scattering with quarks and gluons via exchange of electroweak bosons. In
the limit of pure MDM candidate, the tree level interactions vanish
and the leading interaction occurs via
loops~\cite{Cirelli:2005uq,Hisano:2015rsa}.  Here we mainly discuss
the salient features of the spin independent scattering cross-section on
nucleons at tree level, with particular emphasis on the 5M4D model, while arguing about the expected behavior at loop level. 
A detailed computation of the scattering cross-section in
HMDM should be the subject of a dedicated analysis that is beyond the
scope of this work.

\begin{figure}[t!]
  \begin{center}     
     \includegraphics[width=8.3cm]{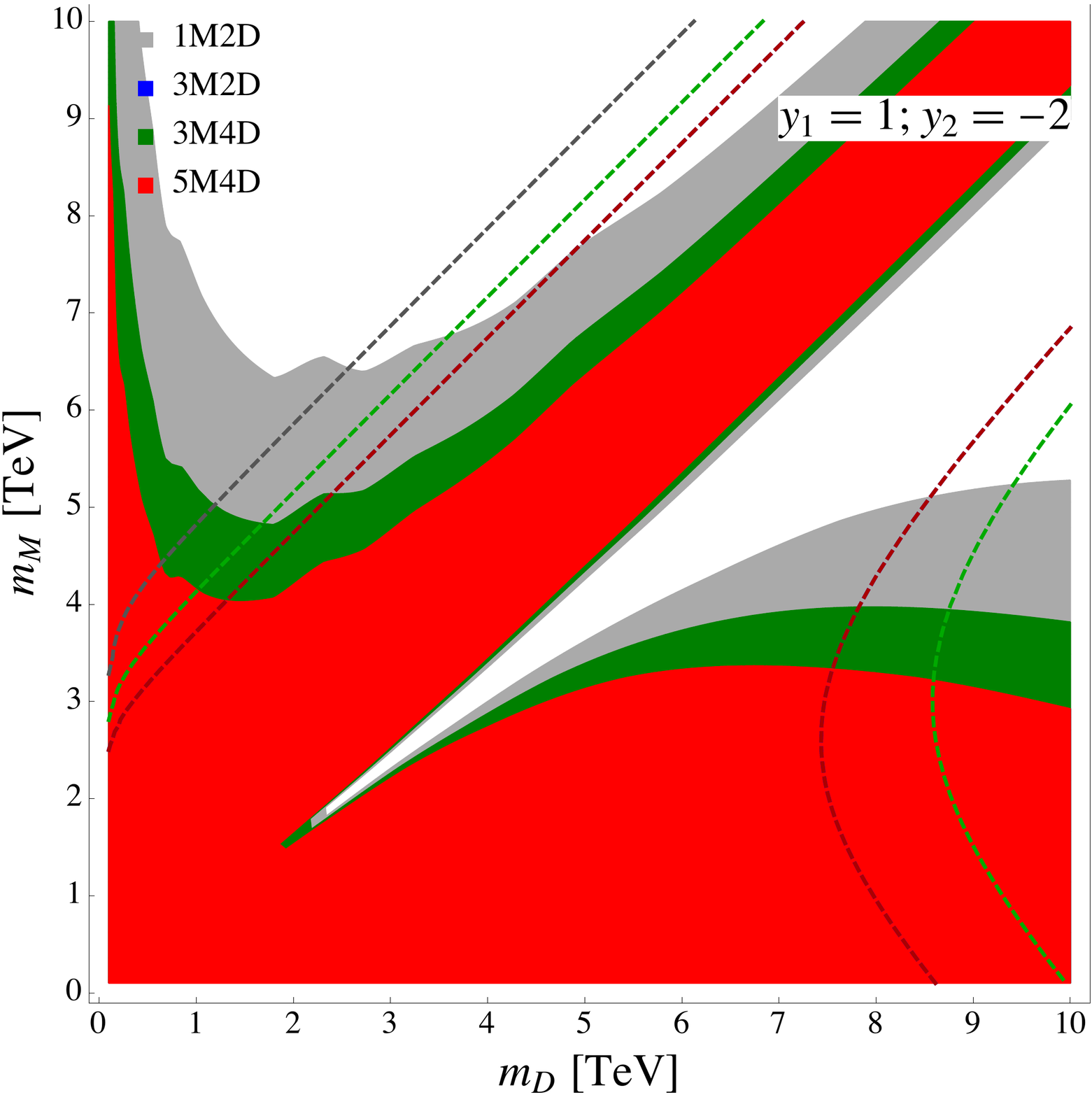}
  \end{center}
  \vspace{-1.2cm}
  \caption{ Tree-level DM-nucleon SI scattering in the plane $m_D-m_M$
    for all the models considered in this work. Colored regions are
    excluded by prospects of limits from the XENON1T experiment~\cite{Aprile:2015uzo} for $y_1=1$ and $y_2 = -2$. Colored lines show
    contours of DM composition, the lower right ones denote 
      $|Z_{11}^M|^2=0.999$ while the upper left ones correspond to
     $|Z_{11}^M|^2=0.001$. The contours of the
     $3_{\textrm{\footnotesize M}}2_{\textrm{\footnotesize D}}$ model
     overlap with those of the $5_{\textrm{\footnotesize
         M}}4_{\textrm{\footnotesize D}}$ model.}
\label{fig:Pante}
\end{figure}

From the discussion in sec.~\ref{sec:neutral-states} focusing on the
custodial symmetry limit, it appears that the DM coupling to the Higgs
(driving the direct detection cross-section at tree-level) is expected
to be {\it maximal} in the limit $m_M\rightarrow m_D$ and
$y_1\rightarrow y_2$ while it is expected to {\it vanish} for
$m_M\rightarrow m_D$ and $y_1\rightarrow -y_2$. Let us see how this
goes beyond the custodial limit.  The SI scattering cross section for
the DM candidate off a nucleon $N$ at tree level for the model $M$ is
\cite{Cheung:2013dua}: \be \sigma^M_{\rm SI} \propto \df{\mu^2}{m_h^4}
(c^M_{h\chi^0_l\chi^0_l})^2~, \ee where $\mu = {m_{\chi^0_l}
  m_N}/{(m_{\chi^0_l} + m_N)}$ is the nucleon-DM reduced mass, $m_h$
is the Higgs mass, and the coefficient $c^M_{h\chi^0_l\chi^0_l}$
contains the Higgs-DM coupling in the model $M$, and is: 
\be
c^M_{h\chi^0_l\chi^0_l} = -\df{c_M}{\sqrt{2}}\left[y_1
  (Z^M_{11})^*(Z^M_{12})^* + y_2 (Z^M_{11})^*(Z^M_{13})^* \right]~.
\ee 
The matrix ${\bf Z}^M$ defines the rotation to the mass basis with
the $\{\chi^0_{\alpha}\}$ states ordered from light to heavy states
($\alpha=l,m,h$). Going from the basis used in
Sec.~\ref{sec:neutral-states}, with $\{\chi^0_{i}\}$ indices $i=1,2,3$
not pointing to any mass ordering, to the basis used here just simply
imply a permutation of the entries of the transformation matrix of
Eq.~(\ref{eq:rot}) in order to get ${\bf Z}^M$.
 Finally, the coefficients $c_M$ for all
the models are: \be c_{\rm 1M2D} = 1,~~~c_{\rm 3M2D} =
\df{1}{\sqrt{2}},~~~c_{\rm 3M4D} = \sqrt{\df{2}{3}},~~~c_{\rm 5M4D} =
\df{1}{\sqrt{2}}~.  \ee

We show in Fig.~\ref{fig:Pante} the present and future exclusion
region from XENON1T experiment~\cite{Aprile:2015uzo,Aprile:2017iyp}
from the calculation at tree-level for a choice of Yukawa couplings
$y_1=1$ and $y_2=-2$. As can be seen, there are common features to all
models considered above. First, there are parts of the parameter space
where the cross section is suppressed, even for light DM that is
largely mixed. In Fig.~\ref{fig:Pante}, this translates as incursions
of the white area into the colored regions illustrating the reach of
Xenon 1T for a given choice of $y_1$ and $y_2$. Around these ``blind
spots", the coupling of the Higgs to DM that mediates the tree-level
interactions is suppressed, as has been discussed in the literature
for the case of the supersymmetric neutralino \cite{Cheung:2012qy} and
the $1_M2_D$ model \cite{Cheung:2013dua,Calibbi:2015nha}.  Second, for
a given size of the Yukawa couplings and for large enough masses the
composition of DM seems to depend on $m_M-m_D$. Indeed, as observed in
\cite{Hill:2013hoa}, in this limit the dynamics can be described in
terms of two parameters only, $\Delta=(m_M-m_D)/ 2$ and
$a=|y_1+y_2|/2$ for real Yukawa couplings. The reason is that the
DM-Higgs effective vertex is in this case proportional to
$a^2/\sqrt{a^2+(\Delta/2 m_W)^2}$. In the $\Delta\to 0$ limit
(i.e. along the diagonal), the cross section is thus maximised. This
behavior generalizes the dependence in $\Delta$ and $a$ that we
observed in the custodial symmetry limit in
Sec.\ref{sec:neutral-states} .

\begin{figure}[t]
  \begin{center}     
    \begin{tabular}{cc}
      \includegraphics[width=7.5cm]{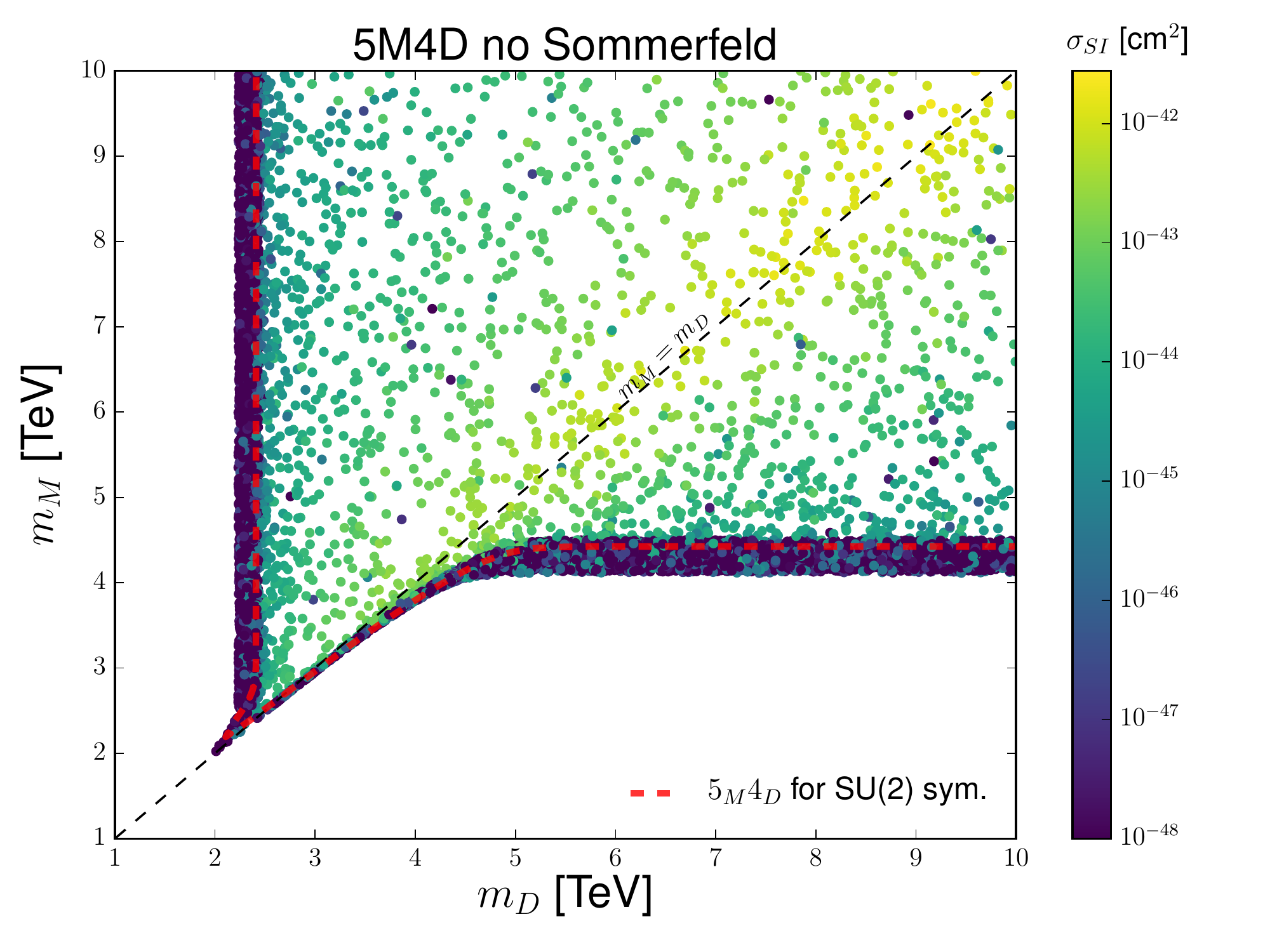}
      \includegraphics[width=7.5cm]{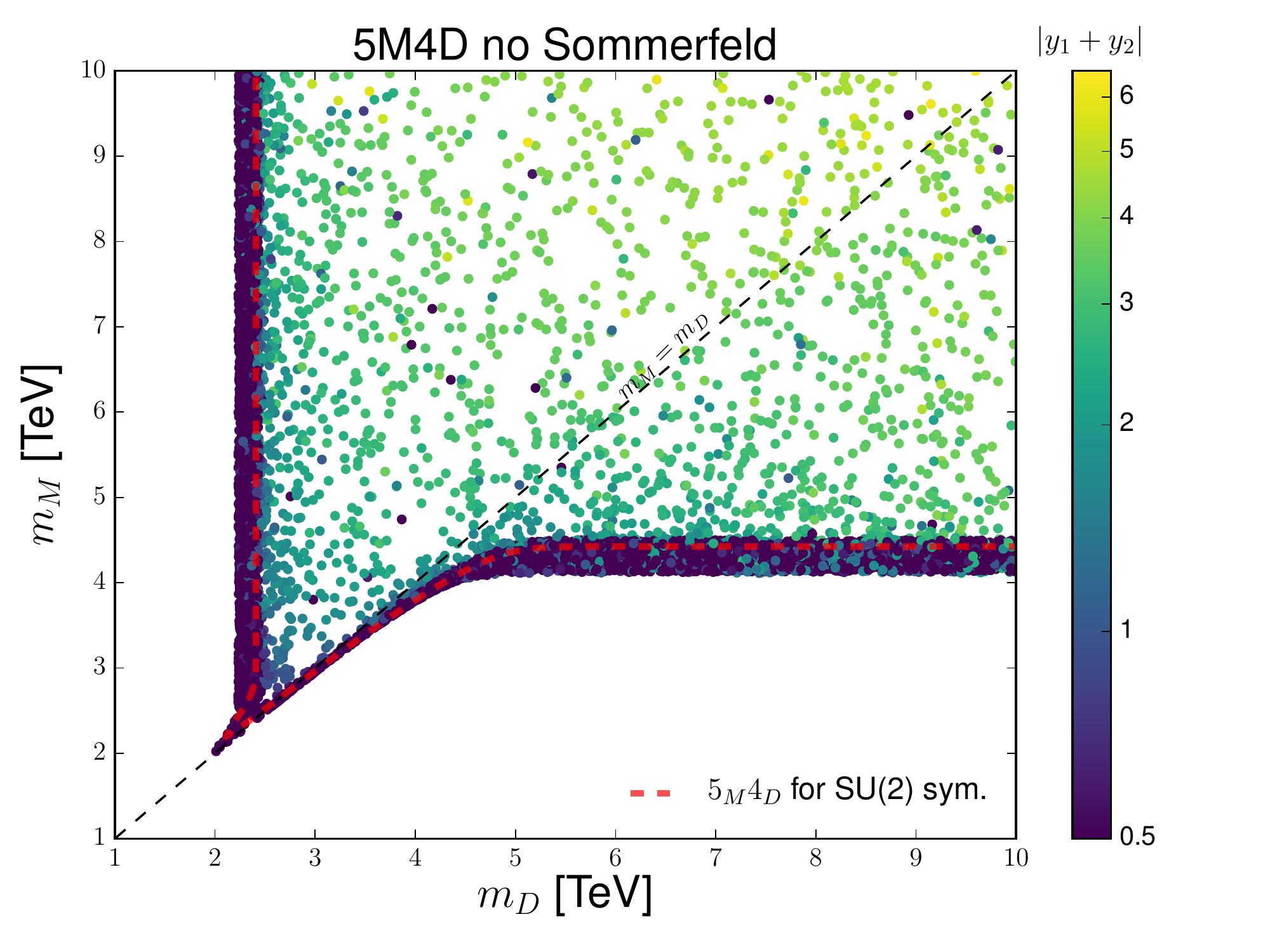}
    \end{tabular}
      \end{center}
  \caption{Viable parameter space in the perturbative 5M4D case for an
    explicit integration of the dark matter abundance with {\tt
      micrOMEGAs} in the $SU(2)_L$-broken limit as in
    Fig.~\ref{fig:pert5M4D-yeff}. All points give rise to $\Omega
    h^2\simeq 0.12$ and the value of the corresponding $\sigma_{SI}$
    and $|y_1+y_2|$ are indicated with the color code in the left and
    right plot respectively. With red dashed line, we show the contour
    obtained in the $SU(2)_L$ symmetric limit for the 5M4D case in
    Fig.~\ref{fig:pert}.}
\label{fig:5M4Dtree}
\end{figure}

Let us now illustrate the above discussion in a concrete HMDM
model. We focus on the 5M4D model for which we have already discussed
the viable parameter space in Sec.~\ref{sec:DM-viab}.  In particular
the results of Fig.~\ref{fig:pert5M4D-yeff} were obtained from a
random scan in the $SU(2)_L$ broken limit with all calculations at
tree-level using {\tt micrOMEGAs}. Here we project in
Fig.~\ref{fig:5M4Dtree} the same parameter space in the $m_M-m_D$
plane with, this time, the gradient color corresponding to the
values of the spin independent scattering cross-section computed with {\tt micrOMEGAs}, $\sigma_{SI}$, on the left hand (LH) side and $|y_1+y_2|$
on the right hand (RH) side. Let us first focus on the LH side plot
illustrating the $\sigma_{SI}$ dependence on the parameters. The largest values of $\sigma_{SI}$ clearly appear to cluster along the diagonal, i.e. $\Delta=0$ as expected from the above discussion. On the other hand, the
dark blue colored points correspond to the vanishing tree-level
$\sigma_{SI}$. Most of them appear to cluster at the boundary of the
viable parameter space, i.e. for vanishing Yukawas or pure MDM
cases. In addition, we see that some more blue points appear to have a
suppressed $\sigma_{SI}$ outside from the boundaries, within the mixed
region.  Comparing the LH side plot to the RH side plot, illustrating
the dependence in $|y_1+y_2|$, it appears that there is clearly a
close correlation between suppressed $\sigma_{SI}$ (darker points on
the RH side) and vanishing $|y_1+y_2|$. In the mixed region, we know
from Fig.~\ref{fig:pert5M4D-yeff} that such points typically have
non-zero $\sqrt{y_1^2+y_2^2}$ values. As a consequence, we can see
that, in the 5M4D case (at tree-level), points with suppressed
$\sigma_{SI}$ and non negligible Yukawa couplings can be obtained
$y_1\rightarrow -y_2$ corresponding to $a\rightarrow 0$ in agreement with the above discussion.

\begin{figure}[t!]
  \begin{center}     
     \includegraphics[width=12cm]{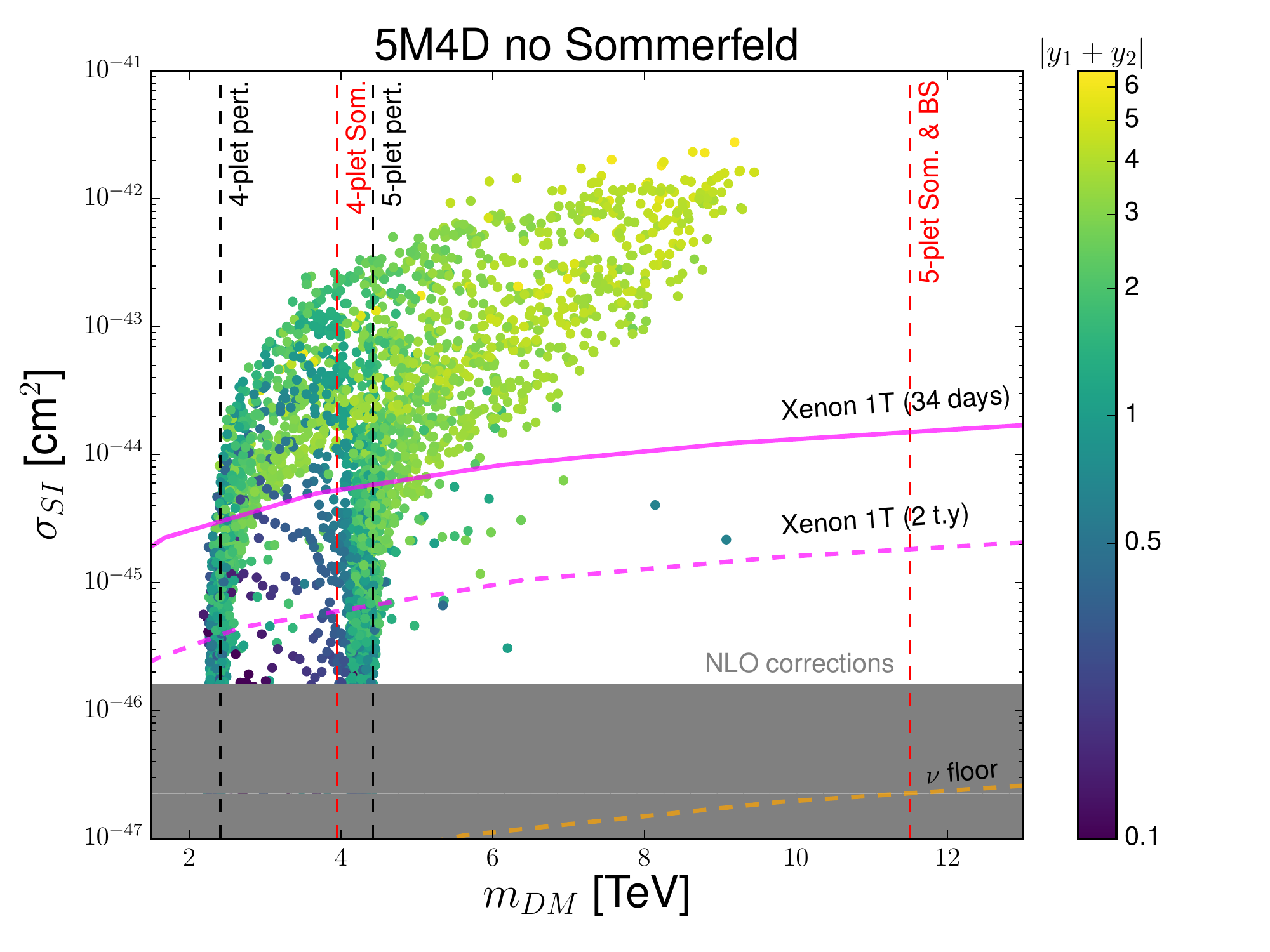}
  \end{center}
  \caption{  $\sigma_{SI}$ in the 5M4D case for an explicit
    integration of the dark matter abundance with {\tt micrOMEGAs} in
    the $SU(2)_L$-broken limit as in Fig.~\ref{fig:pert5M4D-yeff}. All
    points give rise to $\Omega h^2\simeq 0.12$. The gray zone is
    expected to be strongly affected by NLO corrections as in this
    zone $\sigma_{SI}<1.6\,10^{-46}$ cm$^2=\sigma_{SI, {\rm
        4-plet}}^{\rm NLO}$ computations. The vertical black dashed
    lines indicate the DM mass obtained in the $SU(2)_L$ symmetric
    limit for the pure 4-plet and 5-plet case without sommerfeld
    corrections. The red dashed lines include the Sommerfeld
    correction for the 4-plet and the Sommerfeld + Bound state effects
    from~\cite{Mitridate:2017izz} in the 5-plet case. The continuous
    magenta line denote the current constraints from the Xenon 1T
    experiment~\cite{Aprile:2017iyp} and the magenta line shows the
    reach prospects for the same experiment~\cite{Aprile:2015uzo}. The
    dashed orange line shows the ``discovery limit''
    from~\cite{Billard:2013qya}.}
\label{fig:5M4D-SI-M}
\end{figure}

Figure \ref{fig:5M4D-SI-M} shows the same information as the LH plot
of Fig.~\ref{fig:5M4Dtree} but now in the $\sigma_{SI}$ vs. $m_{DM}$
plane, where the color represents the value of $|y_1+y_2|\propto
a$. Again, all the points in the scatter plot reproduce the observed
relic abundance computed without taking into account Sommerfeld nor
bound-state formation. However, we may expect that these corrections will only shift (and enlarge) the
overall shape of the points cloud to the right, and that the features will
remain the same.  Around the pure limits, ie near the vertical dashed
lines without (with) non-perturbative corrections in black (red)
color, the tree-level $\sigma_{SI}$ can typically be much smaller than
for the mixed regions (away from the vertical dashed lines) and even
below the direct detection experiments prospects. In these regions, we
expect that the loop corrections are quite relevant. As a guide for
the eye, we show with gray color in Fig.~\ref{fig:5M4D-SI-M}, the
region where electroweak corrections already appear to be relevant. In
practice we do not expect to have cross-sections, including NLO
corrections, to sum up well below the pure 4-plet result $\sigma_{SI,
  {\rm 4-plet}}^{\rm NLO}= 1.6\,10^{-46}$ cm$^2$ obtained
in~\cite{Hisano:2015rsa}. In practice, Higgs mediated loop corrections
should provide some extra features. Some estimation of this effect is
already provided by~\cite{Hill:2014yka,Hill:2013hoa} for the $1_M2_D$
and the $3_M2_D$ models taking into account two-loop contribution to
the twist-2 gluon effective operator and running of the Wilson
coefficients down to the nuclear scale.\footnote{ Notice that the more
  recent analysis of~\cite{Hisano:2015rsa} took into account extra
  contributions that slightly modify the conclusion
  of~\cite{Hill:2014yka,Hill:2013hoa} for the pure cases. } The main
feature that we underline here also is that the tree level
cross-section dominates in the region of $m_M=m_D$ or equivalently
$\Delta=0$. Beyond tree-level, loop-level blind spots could occur
because of a  cancellation between the contribution from the
scalar and the twist-2 operators\footnote{ New blind spots at loop
  level could appear in intermediate $\Delta$ region  in all cases
  except for the singlet-like limit of the $1_M2_D$ model, see ~\cite{Hill:2014yka,Hill:2013hoa}.}, as shown
e.g. in~\cite{Hill:2014yka,Hill:2013hoa}.

%
%

\subsubsection{Discussion of indirect searches} 
\label{sec:brief-disc-indir}

In section \ref{sec:DM-phen-beyond}, we estimated the impact of the Sommerfeld effect on the relic abundance, which is clearly important in estimating the mass of the thermal candidates. By the same token, the Sommerfeld corrections
can affect DM annihilation in the recent Universe, like at the Galactic Centre,  where the DM is highly non-relativistic. In particular,  they can lead to annihilation cross sections that are much larger (potentially by orders of magnitude) than the canonical value $\sim 3 \cdot 10^{-26}$ cm$^2$/s required for the relic abundance  \cite{Hisano:2004ds}. This is particularly true for large multiplet Minimal Dark Matter candidates, not only because they tend to be in the TeV mass regime, substantially larger than the mass of the $Z$ and $W$ gauge bosons, but also because their multiplet contain particles multiply charged under $U(1)_{\rm em}$. This aspect of MDM has been much studied, starting with~\cite{Cirelli:2007xd} (see also \cite{Cirelli:2009uv}). Calculating the Sommerfeld corrections is  infamously involved because of resonant behaviors due to mass splittings, and the results have been somewhat varying in time (but eventually converged, see Figure 7 \cite{Cirelli:2015bda} and Figure 3 in \cite{Garcia-Cely:2015dda}).\footnote{For similar considerations regarding Wino DM $\equiv$ 3-plet MDM, see e.g.\cite{Cohen:2013ama,Baumgart:2014saa,Beneke:2016ync}. }

 A pure fermionic minimal dark matter candidate is strongly
 constrained by searches for gamma-ray spectral features
 (e.g. monochromatic lines) from the GC region by the HESS
 collaboration~\cite{Ackermann:2013yva}. The 3-plet and the 5-plet are
 both are excluded if the DM profile is cuspy, NFW or Einasto, while
 the 5-plet is marginally viable if the profile is cored, isothermal
 or Burkert
 \cite{Cohen:2013ama,Hryczuk:2014hpa,Cirelli:2015bda,Garcia-Cely:2015dda,Ovanesyan:2016vkk,Lefranc:2016fgn}.~\footnote{Notice
   that a priori one could also get monochromatic photon emission from
   bound state ($B$) formation processes $\chi_0 \chi_0\rightarrow
   B\gamma$. For the pure 5-plet case, the latter gamma ray signal
   (with $E_\gamma\ll m_{DM}$) appear to be below the current
   Fermi-LAT telescope sensitivity but could potentially be tested in
   the future depending on the DM mass, see~\cite{Mitridate:2017izz}
   for more details. } Does mixing of a Majorana multiplet with two
 Weyl states bring anything new?  To fully address this question one
 should calculate the non-perturbative corrections for each possible
 viable candidate, taking into account mixing and also the existence
 of new channels associated to Higgs exchange, etc. This is a very
 technical task, way beyond our scope. Instead we merely argue that,
 if anything, mixing brings some new freedom, possibly relaxing the
 constraints from gamma-rays observations. The key point is basic, and
 has been partly considered in some works for the case of Minimal Dark
 Matter candidates, either to enhance or deplete the annihilation
 cross sections at low velocities, see
 e.g. \cite{Chun:2012yt,Chun:2016cnm}.\footnote{More extensive and
   in-depth analyses have been done in the case of the Higgsino-Wino
   mixing, related to search for supersymmetric DM candidates
   \cite{Beneke:2016jpw}. Given the know-how \cite{Beneke:2014gja}, it
   could be interesting to extend such analysis to higher multiplets.}
 It rest on the fact that Sommerfeld corrections that lead to
 mono-chromatic gamma-rays are very sensitive to the mass splitting
 between the DM candidate and its charged partners.  For pure MDM
 candidates, the splitting is set by electroweak corrections, while
 mixed states receive an extra contribution from their direct coupling
 to the Higgs.  Simple criteria to assess the impact of mass splitting
 on the Sommerfeld corrections are given in \cite{Slatyer:2009vg}.
 Suppressing the effect of excited states requires that the mass
 splitting $\Delta$ is larger than the kinetic energy of the DM,
 $m_{\rm DM} v^2/2 \leq \sim \Delta m$. Less obvious, but natural, it
 that the binding energy of DM in an attractive channel, $\sim
 \alpha^2 m_{\rm DM}$ must be smaller than the energy required to
 produced an excited state, $\Delta m$. Regardless, changing $\Delta
 m$ allows to move around the position of the resonant peaks, as is
 for instance illustrated in \cite{Chun:2012yt} and can potentially
 help in evading the gamma-ray constraints.

\section{Conclusion}
\label{sec:concl-prosp}

In the Minimal Dark Matter framework, a dark matter candidate is the neutral component of an electroweak multiplet of dimension $n$. As such a candidate has only gauge interactions,  all observables are in principle univocally determined. In particular, its relic abundance through thermal freeze-out can match the cosmological observed value only for a unique dark matter mass. Also, their signal in both direct and indirect searches are fixed, at least modulo astrophysical uncertainties. As such, they are very useful benchmark WIMP candidates. Focusing on fermionic cases, the highest possible representation, at least  if ones want to avoid Landau poles at low energies, is a Majorana 5-plet. A nice feature of such candidate is that it may be automatically long-lived, without the need of imposing some symmetry, as its coupling to SM degrees of freedom can only come through a dimension 6 operator. Lower dimension representations are nevertheless of much interest, if anything because they correspond to specific corners of well-motivated candidates. For instance, a Majorana triplet is equivalent to a pure wino candidate, while a doublet is a pure higgsino. The latter has non-zero hypercharge, and so is excluded by direct detection if it is a pure Dirac state but mixing with a triplet or a singlet (i.e. a bino), through the Higgs doublet, makes it Majorana (or quasi-Dirac). 

In this work we have extended on the Minimal Dark Matter framework by considering all pairs of electroweak fermionic multiplets (up to a 5-plet) that can have a Yukawa coupling with  the Standard Model Higgs doublet, a framework we dubbed Higgs coupled Minimal Dark Matter or HMDM. 
 As in the MDM framework, avoiding the
Landau pole for the EW coupling at a low scales, 
we end up considering four possible models of mixed Majorana and Dirac fermions, including the $1_M2_D, 3_M2_D, 3_M4_D$ and $5_M4_D$.  Because of mixing, and the coupling to the Higgs, the phenomenology of such scenarios is much more involved than in the pure MDM case. Several cases have been already considered in the literature, in particular in relation with the neutralino candidates to which we alluded to above. The $3_M 4_D$ case has only been discussed recently, see \cite{Tait:2016qbg}. To our knowledge, the $5_M 4_D$ case the has not yet been considered in the literature.

Our purpose was to provide a unified presentation of the different cases. Doing so, we have first provided a detailed analysis of the
dark matter mass spectrum. We have  made use of the existence of a custodial symmetry that arises for specific Yukawa couplings and that provides a way to understand many features of the mass spectra, including the emergence of  quasi-degenerate electroweak multiplets and an understanding of the mass splitting between the components. In particular, we have shown that, at tree level, the lightest neutral particle (LNP) is always the lightest component, and so potentially a dark matter candidate. This conclusion has however to be moderated as one-loop corrections may change the hierarchy of masses, a fact that we have inferred from \cite{Tait:2016qbg} and their analysis of the $3_M 4_D$ case. Next, we have then analyzed the viable parameter space of HMDM both in the
perturbative approximation and taking into account non-perturbative effects. Indeed, as is the case of MDM,
the candidates considered here are expected to be particularly affected by
Sommerfeld effect and also, in the case of largest $SU(2)_L$ representations, by
bound state formation. The calculations of these phenomena is notoriously delicate, and even more so for mixed candidates, and have only been tackled for specific  mixed scenarios associated to SUSY phenomenology.  Here, we have merely 
extracted the boundaries of the viable HMDM parameter space, and this using the electroweak symmetric limit,  both for the perturbative regime and for
non-perturbative corrections. This procedure greatly simplifies the
calculations and yet, we argued, provides a good proxy to more precise calculations. Doing so, we have provided the first estimate of the mass of a
(quasi-pure) 4-plet candidate, taking into the Sommerfeld
effects. Figure~\ref{fig:som} and Tab.~\ref{tab:svI} summarize our
findings for all the considered HMDM scenarios.  

The HMDM framework greatly increases the range of possible DM candidates. Their coupling to the Higgs, on top of gauge bosons, also greatly enhances the possibility for their search through direct detection experiments. This is clear using the parameter space of HMDM candidates using only perturbative calculations. We have argue that the same should hold taking into account the correction on the mass of the dark matter candidates due to Sommerfeld effect. In particular, several candidate in the multi-TeV range should be within reach of the current Xenon-1T experiment and, a fortiori, of future direct detection experiments. 
We have not addressed in details indirect detection, for which Sommerfeld corrections are particularly at the same time very relevant and very sensitive to the precise characteristics of not only the LNP particle, but also of the other components of the electroweak multiplet to which it may belong, and in particular the mass splittings, which in the HMDM scenario arises at tree level, except at exceptional custodial points. A complete analysis would require to take into account a full one-loop calculation of the mass spectrum, as well as the Sommerfeld effects. Such study remains to be done for the $3_M4_D$ and $5_M 4_D$ cases, which are of particular interest as they point to DM candidate in the multi-TeV mass range. We leave this however for future works.

\section*{Acknowledgments}

We thank C.~Garcia-Cely and T.~Slatyer  for helpfull discussions. L.L.H. and M.T. are 
supported by the FNRS-FRS, the Belgian Federal Science Policy
Office through the Interuniversity Attraction Pole P7/37 and the IISN.
L.L.H. is also supported by the Strategic Research Program High Energy Physics and the
Research Council of the Vrije Universiteit Brussel. B.Z. is supported by the {\it Investissements d'avenir} Labex ENIGMASS.

\appendix

\section{Generators of $SU(2)$ and other useful formulas}
\label{sec:tensor-formalism}

We enlist all the generators of the su(2) algebra up to the 6-dimensional representation
{\footnotesize\bea
&&T_2^1=\left(\begin{array}{cc}
0 & {1\over  2} \\
{1\over  2} & 0
\end{array}\right)\,,\quad
T_2^2=\left(\begin{array}{cc}
0 & -{i\over  2} \\
{i\over  2} & 0
\end{array}\right)\,,\quad
T_2^3=\left(\begin{array}{cc}
{1\over  2} & 0 \\
0 & -{1\over  2}
\end{array}\right)\,,\nn
\\
&&T_3^1=\left(\begin{array}{ccc}
0 & {1\over  \sqrt{2}} & 0 \\
{1\over  \sqrt{2}} & 0 & {1\over  \sqrt{2}} \\
0 &  {1\over  \sqrt{2}} & 0
\end{array}\right)\,,\quad
T_3^2=\left(\begin{array}{ccc}
0 & {-i\over  \sqrt{2}} & 0 \\
{i\over  \sqrt{2}} & 0 & {-i\over  \sqrt{2}} \\
0 &  {i\over  \sqrt{2}} & 0
\end{array}\right)\,,\quad
T_3^3=\left(\begin{array}{ccc}
1 & 0 & 0 \\
0 & 0 & 0 \\
0 &  0 & -1
\end{array}\right)\,,\nn
\\
&&T^1_4\!=\!\left(\!\!\begin{array}{cccc}0 & {\sqrt{3}\over 2} & 0 & 0 \\{\sqrt{3}\over 2} & 0 & 1 & 0 \\0 & 1 & 0 & {\sqrt{3}\over 2} \\0 & 0 & {\sqrt{3}\over 2} & 0\end{array}\!\!\right)\,,\
T^2_4\!=\!\left(\!\!\begin{array}{cccc}0 & \textrm{-}i{\sqrt{3}\over 2} & 0 & 0 \\i{\sqrt{3}\over 2} & 0 & \textrm{-}i & 0 \\0 & i & 0 & \textrm{-}i{\sqrt{3}\over 2} \\0 & 0 & i{\sqrt{3}\over 2} & 0\end{array}\!\!\right)\,,\
T^3_4\!=\!\left(\!\!\begin{array}{cccc}{3\over 2} & 0 & 0 & 0 \\0 & {1\over 2} & 0 & 0 \\0 & 0 & \textrm{-}{1\over 2} & 0 \\0 & 0 & 0 & \textrm{-}{3\over 2}\end{array}\!\!\right)\,,\nn
\\
&&T^1_5\!=\!\left(\!\!\begin{array}{ccccc}0 & 1 & 0 & 0 & 0\\1 & 0 & {\sqrt{6}\over 2} & 0 & 0\\0 & {\sqrt{6}\over 2} & 0 & {\sqrt{6}\over 2} & 0\\0 & 0 & {\sqrt{6}\over 2} & 0 & 1\\0 & 0 & 0 & 1 & 0\end{array}\!\!\right)\,,\ 
T^2_5\!=\!\left(\!\!\begin{array}{ccccc}0 & \textrm{-}i & 0 & 0 & 0\\i & 0 & \textrm{-}i{\sqrt{6}\over 2} & 0 & 0\\0 & i{\sqrt{6}\over 2} & 0 & \textrm{-}i{\sqrt{6}\over 2} & 0\\0 & 0 & i{\sqrt{6}\over 2} & 0 & \textrm{-}i\\0 & 0 & 0 & i & 0\end{array}\!\!\right)\,,\
T^3_5\!=\!\left(\!\!\begin{array}{ccccc}2 & 0 & 0 & 0 & 0\\0 & 1 & 0 & 0 & 0\\0 & 0 & 0 & 0 & 0\\0 & 0 & 0 & \textrm{-}1 & 0\\0 & 0 & 0 & 0 & \textrm{-}2\end{array}\!\!\right)\,,\nn
\\
&&T_6^1\!=\!\left(\!\!\begin{array}{cccccc}
0 & {\sqrt{5}\over 2} & 0 & 0 & 0 & 0 \\
{\sqrt{5}\over 2} & 0 & \sqrt{2} & 0 & 0 & 0 \\
0 & \sqrt{2} & 0 & {3\over 2} & 0 & 0 \\
0 & 0 & {3\over 2} & 0 & \sqrt{2} & 0 \\
0 & 0 & 0 & \sqrt{2} & 0 & {\sqrt{5}\over 2} \\
0 & 0 & 0 & 0 & {\sqrt{5}\over 2} & 0
\end{array}\!\!\right)\,,\ 
T_6^2\!=\!\left(\!\!\begin{array}{cccccc}
0 & -i{\sqrt{5}\over 2} & 0 & 0 & 0 & 0 \\
i{\sqrt{5}\over 2} & 0 & -i\sqrt{2} & 0 & 0 & 0 \\
0 & i\sqrt{2} & 0 & -i{3\over 2} & 0 & 0 \\
0 & 0 & i{3\over 2} & 0 & -i\sqrt{2} & 0 \\
0 & 0 & 0 & i\sqrt{2} & 0 & -i{\sqrt{5}\over 2} \\
0 & 0 & 0 & 0 & i{\sqrt{5}\over 2} & 0
\end{array}\!\!\right),\nn
\\
&&T_6^3=\left(\begin{array}{cccccc}
{5\over 2} & 0 & 0 & 0 & 0 & 0 \\
0 & {3\over 2} & 0 & 0 & 0 & 0 \\
0 & 0 & {1\over 2} & 0 & 0 & 0 \\
0 & 0 & 0 & -{1\over 2} & 0 & 0 \\
0 & 0 & 0 & 0 & -{3\over 2} & 0 \\
0 & 0 & 0 & 0 & 0 & -{5\over 2}
\end{array}\right)\,. \nn
\eea
}


We use the tensor formalism where
\begin{eqnarray}
\left(\!\begin{array}{c}\chi_{1111} \\\sqrt{4}\chi_{1112} \\\sqrt{6}\chi_{1122} \\\sqrt{4}\chi_{1222} \\\chi_{2222}\end{array}\!\right)\equiv\left(\! \begin{array}{c}\chi^{++} \\\chi^{+} \\\chi^{0} \\\chi^{-} \\\chi^{--}\end{array}\!\right)
& \qquad &
\left(\!\begin{array}{c}\psi_{111} \\\sqrt{3}\psi_{112} \\\sqrt{3}\psi_{122} \\\psi_{222}\end{array}\!\right)\equiv\left(\!\begin{array}{c}\psi^{++} \\\psi^{+} \\\psi^{0} \\\psi^{-}\end{array}\!\right) \nonumber\\
& & \\
\left(\! \begin{array}{c}\tilde{\psi}_{111} \\\sqrt{3}\tilde{\psi}_{112} \\\sqrt{3}\tilde{\psi}_{122} \\\tilde{\psi}_{222}\end{array}\! \right)\equiv\left(\! \begin{array}{c}\tilde{\psi}^{+} \\\tilde{\psi}^{0} \\\tilde{\psi}^{-} \\\tilde{\psi}^{--}\end{array}\!\!\!\right)&\qquad &
\left(\! \begin{array}{c}\chi_{11} \\\sqrt{2}\chi_{12} \\\chi_{22}\end{array}\! \right)\equiv\left(\! \begin{array}{c}\chi^{+} \\\chi^{0} \\\chi^{-} \end{array}\! \right)\nonumber
\end{eqnarray}
These normalization factors appearing above just simply correspond to
$\sqrt{{\rm Binomial}[n-1, i - 1]}$, where $n$ is the length of the
multiplet and $i$ is the position of the component of charge $Q$ in
the $T_3$ basis. For the $a_Q$ coefficients defined in (\ref{eq:aQ}),
we have thus for e.g. the neutral component of the Majorana triplet
$a_{\chi^0} =\sqrt{{\rm Binomial}[2, 1]}=\sqrt{2}$ while for the
neutral component of the Majorana quintuplet we have $a_{\chi^0}
=\sqrt{{\rm Binomial}[4, 2]}=\sqrt{6}$.

\section{Cross-sections for two-particle states in $SU(2)$ symmetric limit}
\label{sec:svij2svI}

On can recast the cross-sections $\sigma v_{ij}$, where $ij$
characterizes the two initial state particles, in terms of the $\sigma
v_I$ associated to eigenstates of total isospin $I$ in the $SU(2)_L$
symmetric limit. For the latter purpose, one has to derive the
coefficients $C_{I_a,ij}$ relating a total isospin 2 particle states
$|I_a \rangle$ to a sum states $|ij\rangle$.  This is obtained
inverting the Clebsch-Gordan decomposition of $|ij\rangle$ in terms of
$|I_a \rangle$.\footnote{To exctact the Clebsch-Gordan coefficients,
  $i$ and $j$ can be tagged by their isospin projection (or
  equivalently their charge when $Y$=0) associated to initial
  particles.} The relation between cross-sections then reads:
\begin{equation}
  \sigma v_{ij}^{\rm pert}=\sum_I |C_{I,ij}|^2\sigma v_{I}^{\rm pert}
\label{eq:svij2I}
\end{equation}
where $I$ runs over the $I_a$ values with $a=1,..,N'$. In our case, we
have obtained the analystica expressions of $\sigma v_{ij}$ making use
of {\tt Calchep}.

Below, we detail the derivation of the different contributions to the
total annihilation cross-section in the case of the quadruplet. Notice
that we have provided the relevant $\sigma v_{I}^{\rm pert}$ for all
cases of interest for this paper in Tab.~\ref{tab:svI}. In the
quadruplet case, one considers the annihilation of a $4$ with a $\bar
4$ with hypercharges $Y_4= 1/2$ and $Y_{\bar4}= -1/2$ and
respectively. The index $i$ in $\sigma v_{ij}$ denotes the charge of
annihilating component of the 4 while the index $j$ denotes the charge
of annihilating component of the $\bar 4$. The $SU(2)_L$ only
contributions to the annihilation cross are given by:
\begin{itemize}
\item $Q_{tot}=i+j=0$
\begin{eqnarray}
\sigma v_{0,0}&=& \sigma v_{I=0}/4+\sigma v_{I=1}/20+\sigma v_{I=2}/4= \sigma v_{+,-} \label{eq:sv0b0}\\ 
\sigma v_{++,--}&=& \sigma v_{I=0}/4+9\sigma v_{I=1}/20+\sigma v_{I=2}/4= \sigma v_{-,+} \label{eq:sv2m2}
\end{eqnarray}
\item $Q_{tot}=i+j=1$
\begin{eqnarray}
 \sigma v_{++,-}&=&3\sigma v_{I=1}/10+\sigma v_{I=2}/2= \sigma v_{0,+}\\
 \sigma v_{+,0}&=& 2/5\sigma v_{I=1} \label{eq:sv1b0}
\end{eqnarray}
\item $Q_{tot}=i+j=2$
\begin{eqnarray}
 \sigma v_{++,-}&=&\sigma v_{I=2}/2= \sigma v_{+,+}\,.
\end{eqnarray}
\end{itemize}
Notice that in this case $\sigma v_{ij}\neq \sigma v_{ji}$ as the charge indices
are not the good representative quantum numbers to specify the isospin
projection of each of the annihilating particles that have opposite 
hypercharges.

Using eq.~(\ref{eq:om}), with $\zeta=1/2$ for a Dirac dark matter
particle, the relic abundance can be computed using
\begin{eqnarray} 
\sigma v_{eff} = \sum_{ij} \df{g_ig_j}{g^2_{tot}}\sigma v_{ij} = \df{1}{16}&(&\sigma v_{0,0} + \sigma v_{+,-} + \sigma v_{-,+} + \sigma v_{++,--} 
+ 2(\sigma v_{+,0} + \sigma v_{0,+}) + 2\sigma v_{++,-}
\nonumber\\ 
&+& 2\sigma v_{+,+} + 2\sigma v_{++,0} + 2\sigma v_{++,+})
\label{eq:sveff4Q}
\end{eqnarray}
where the index $i$ and $j$ of the annihilation cross-section $\sigma
v_{ij}$ refer here to the charges of $\psi$ and $\tilde\psi$
respectively.
Using the Clebsh-Gordan decomposition one can extract the
$\sigma v_{I}$, with $I=0,1,2,3$, from the $SU(2)_L$ contributions to
$\sigma v_{ij}$, i.e. the non zero contributions for $g'\rightarrow
0$.\footnote{Here,
  $\sigma v_{++,+} = \sigma v_{I_{\cal R}=7} = 0$, since there is no
  2-particle SM final state with $I>2$.} The expression of $\sigma
v_{eff}$ can then be rewritten as:
\begin{equation} 
\sigma v_{eff} = \df{1}{16}\left(\sigma v_{I=0} + 3\sigma v_{I=1} + 5\sigma v_{I=5} + \sigma v_{g'} + \sigma v_{g'g}\right)
\label{sigmavI}
\end{equation}  
with $\sigma v_{g'}$ and $\sigma v_{gg'}$ being the $U(1)_Y$ and mixed
$U(1)_Y$ \& $SU(2)_L$ contribution as in eq.~(\ref{eq:sveffSIY}).
In the s-wave limit, we have thus found for the 4-plet the results of eq.~(\ref{eq:sv4-plet}).

\bibliography{GSMDM-2.bib}{} 

\bibliographystyle{hunsrt}
\end{document}